# Contribution of shear strains to the vibrational entropy of defective configurations

J.L. BOCQUET*

Service de Recherches de Métallurgie Physique, CEA Saclay,
F-91191 Gif sur Yvette Cedex, France

Calculating the vibrational entropy of an N-atom assembly in the harmonic approximation requires the diagonalisation of a large matrix. This operation becomes rapidly time consuming when increasing the dimensions of the simulation cell. In the studies of point defects, a widely used shortcut consists in calculating the eigen modes of the atoms contained in an inner region, called the defective region, while the atoms belonging to the outer region are held fixed, and in applying an elastic correction to account for the entropy stored in the distortion of the outer region. A recent paper proposed to base the correction on the local pressure change experienced by each lattice site. The present contribution is an extension in the sense that it includes the shears. We compared the two approximations for configurations which are currently encountered in defect studies, namely those pertaining to defect formation and migration. The studied defects are the single, di- and tri-vacancy as well as the dumbbell interstitial in a host matrix modelled by several empirical potentials mimicking pure copper. It is shown that the inclusion of shears brings a noticeable contribution to the elastic correction for all configurations of low symmetry.

## 1. Introduction

The calculation of free energy is of great concern in all the fields of materials science. If the evaluation of potential energy has become today a routine task, the evaluation of the entropy remains a difficult one. Ignoring hereafter the configurational and electronic parts of the entropy, the present work will focus on the vibrational contribution in pure substances with or without point defects.

Modern computing facilities have promoted Monte-Carlo techniques: they are today considered as the best candidates for quantitative determinations of free-energy or free-energy differences in systems, while taking fully into account the detail of the atomic interactions as well as the boundary conditions [1]. Being stochastic in nature, they obtain a result only at the end of a long series of acceptations and rejections; the convergence towards the aimed at quantity with an acceptable uncertainty is never a trivial question and implies a very heavy computational burden. As a consequence their use is most often restricted to those cases where the determination of some basic quantity with a very high precision is highly desirable, the latter playing later on the role of a reference value against which all subsequent determinations or approximations will be rated.

These highly sophisticated techniques cannot be used routinely whenever a large number of independent calculations is involved in the physical phenomenon to be modelled, like for example diffusive phase transformations described by successive atomic jumps with on-the-fly determinations of the diffusional parameters [2, 3, 4]. This is the reason why alternate and more tractable routes involving less calculations have constantly been looked for, even if the price to be paid for this is, undisputably, some loss of accuracy. In this spirit, the efficiency of harmonic and quasi-harmonic approximations rests on their ability to reduce the initial problem to a deterministic calculation manipulating algebraic objects. They have been extensively explored during decades for studying the formation and the migration of point defects, in metallic alloys [5] as well as ionic crystals [6] and have proved their relevance by obtaining in many cases the right order of magnitude for these basic quantities which are accessible to an experimental determination.

At the light of the most recent results, one begins to understand the basic reasons why, and the conditions under which, the harmonic (or quasiharmonic) approximation is expected to work. The latter focusses on the curvatures of an harmonic basin of potential energy around some equilibrium state. The situations where this approach fails and departs substantially from more exact evaluations are encountered whenever the temperature raise leads to the emergence of several metastable minima, the energy of which are comparable to the one of the primitive harmonic basin: this case is typically encountered in amorphous systems and metallic nano-aggregates. For specific model systems, all these minima can be detected exhaustively with their accurate Boltzmann weight through the use of a convenient Monte-Carlo sampling procedure [7]. Then the harmonic superposition approximation taking into account all these minima with their thermodynamical weights does yield a very accurate description of the free eneergy over a very large temperature range and predicts transition temperatures[1] with an excellent precision [8]. Coming back to point defect studies, the basic reason why the harmonic or quasiharmonic approximations are so successful is linked to a particularly favourable circumstance: the introduction of a point defect in a pure matrix is not a perturbation of sufficient extent to question the unicity of the harmonic basin describing the undefective reference state and to lift its degeneracy. As a consequence, following the shift of the eigen modes of this unique basin is sufficient to have a good representation of the free energy changes versus defect introduction and temperature raise. In all what follows we will confine ourselves to the most simple harmonic approximation.

In this frame, the vibrational entropy of a solid containing N atoms is evaluated as the entropy of N coupled three-dimensional harmonic oscillators and reads [9] :

$$S = k \sum_{i=1}^{3N} \left[ x_i (\exp\{x_i\} - 1)^{-1} - \text{Log}(1 - \exp\{-x_i\}) \right] \qquad (1)$$

where $x_i = \hbar\omega_i / kT$ and the $\omega_i$ are the eigen angular frequencies of the system. In the high temperature approximation, the expression reduces to :

$$S \simeq k \sum_{i=1}^{3N} \left[ 1 - \text{Log}\{x_i\} \right]. \qquad (2)$$

The normal modes of this assembly are calculated with the use of the force-constant matrix **D**, containing the second derivatives of the potential energy **U** versus the displacements of atoms 'i' and 'j' of masses $m_i$ and $m_j$ along the directions 'α' and 'β' respectively and expressed by:

$$D_{ij}^{\alpha\beta} = \left[ m_i m_j \right]^{-1/2} \frac{\partial^2 U}{\partial x_i^\alpha \partial x_j^\beta}. \qquad (3)$$

### 1.1. *Supercell method*

The supercell method consists in applying the above equations to a periodically repeated crystallite of a perfect or defective lattice. The frequencies $\omega_i$ are the phonon modes at zero wave vector in the reduced Brillouin zone corresponding to the large unit cell: three modes have a zero frequency and correspond to the bodily translation induced by the repetition of the simulation cell. After having dropped these three translational modes by deleting the last three rows and columns of **D**, its determinant $\Delta$ is equal to the squared product of the angular frequencies $\omega_i$. In the high temperature approximation, the entropy change is then reduced to:

$$S = -\frac{k}{2}\text{Log}(\Delta) + 3k(N-1)\left[1 - \text{Log}(\frac{\hbar}{kT})\right] . \qquad (4)$$

* Email : Jean-Louis.Bocquet@cea.fr

Any modification of the force constants, stemming from a spatial distortion and/or a mass disorder induced by he presence of a point-or-extended defect, produces a corresponding entropy change :

$$S' - S = -\frac{k}{2}\text{Log}(\Delta'/\Delta) + 3k(N'-N)\left[1 - \text{Log}(\frac{\hbar}{kT})\right] \qquad (5)$$

where the primes denote the defective system. The last term serves to correct the dimensions of the first term whenever the number N' of atoms in the defective state is not equal to that in the reference state, a situation which prevails in all the point defect calculations operating with a constant number of sites.

At last, if the reference state is the perfect crystal, the dimensional correction is commonly carried back onto the perfect contribution by rescaling the number of atoms in the reference state to N':

$$S' - S \simeq -\frac{k}{2}\text{Log}(\Delta'/\Delta^{N'/N}) . \qquad (6)$$

The calculation of the normal modes implies the diagonalization of the force constant matrix **D**. The latter is often introduced as a (3N × 3N) matrix, which is, strictly speaking exact only for those systems with longe-ranged forces like ionic materials. In systems governed by short-ranged forces indeed, the force constants are different from zero only if atoms 'i' and 'j' are apart from each other for less than a distance $d_{vib}$ equal to $r_{cut}$ for pairwise interactions and 2 $r_{cut}$ for N-body interactions, where $r_{cut}$ stands for the interaction range. The number of non-null terms in each line of **D** is thus bounded by the number $N_{vib}$ of neigbours whithin a radius $d_{vib}$. The necessary amount of storage scales then as (3N × 3$N_{vib}$) rather than quadratically versus N. Notwithstanding this remark, the diagonalization of large dynamic matrices, even for systems with short-ranged forces, remains associated with a large computational cost and to numerical problems, as soon as the simulation cell contains more than a few thousands of atoms. This is the reason why, during the last decades, numerous attempts have continuously searched for various kinds of approximations. The supercell method is however often kept as a standard reference point for other approximations because its results are found to vary only weakly with the cell size [10].

### 1.2. *Local approximations of the force constant matrix*

These approximations try to reduce the size of the above matrix on the basis of physical arguments: in metals, the interactions are short-ranged thanks to the electronic screening and a small number of neighbour shells is expected to reproduce the main contribution to the energy and the entropy.

**1.2.1** *Einstein approximation and local harmonic models.* The lowest order approximation (or Einstein approximation) calculates the eigen modes of an assembly of uncoupled particles, where each atom vibrates in the potential well created by its immobile neighbours; the problem is thus reduced to the diagonalisation of (N-1) (3×3) matrices. Increasing gradually the coupling between the central atom and its neighbours by taking into account more and more distant shells provides a complete hierarchy of local harmonic approximations ending up to the supercell approximation. In particular, the approach provides an interesting expression on the determinant Δ which embodies the effects of off-diagonal terms. This route was been followed for calculating defect formation entropies [11, 12, 13] as well as pre-exponential factors for migration [2]. Although interesting for their pedagogical content, these approximations are however not satisfying: they predict a correct variation of the free energy as a function of temperature for a perfect lattice; but in most cases, the mere introduction of a vacancy spoils their success because their inherent simplicity precludes completely capturing the quasilocalized modes around the point defect and offering truly quantitative evaluations of formation entropies.

**1.2.2. *Second moment approximation for the local density of states.*** This approximation postulates a reasonable analytical form for the vibrational local density of states which rests on a second moment approximation. A simple analytical formulation can be obtained for the definition of temperature dependent atomic forces, which opens the route towards the determination of temperature dependent equilibrium structures through the direct minimization of a free energy. This formulation was used for the calculation of grain boundary structures as a function of temperature and was able to determine those sites which brought the largest contributions to the excess entropy. The second moment approximation has however several drawbacks: i) it requires that the potential has an analytical formulation of its third derivative continuous at all interatomic separations, a condition which is not fulfilled by many currently available potentials; ii) it does not take into account the off-diagonal elements of the (3 ×3) force constant matrix and obtains a free energy which is less accurate than the one provided by the local harmonic formulations quoted above; iii) it can take properly into account the atomic arrangements involving local pressure only but not shears; going further would necessarily imply the introduction of fourth order moments, at the expense of considerably greater complexity [14, 15].

**1.3. *Embedded crystallite method***

The method consists in partitioning the defective crystal into two regions: an inner region containing N* defective site(s) with a prescribed set of neighbours (often called the embedded crystallite) and an outer one containing the N-N* remaining atoms. Two routes have been followed then.

**1.3.1. *Green function method.*** This route is semi-analytical and rests on the use of a Green function method. The entropy formula is recast in terms of the Green functions of the perfect lattice and the changes of the force constants caused by the introduction of the defect. The main assumption is that the force constants are changed only in the inner region. Several treatments can be done according to whether the atomic positions used for the calculation of force constants are the perfect positions, or the fully relaxed ones, thus producing results of increasing accuracy [16]. In ionic crystals, the problem is further complicated by several features which are specific to ionic bonding: long-ranged forces, alternate charges of successive shells [17]. The pedagogical interest of Green function formulations is obvious since tractable analytical expressions can be obtained under simplifying assumptions; however, quantitative results are obtained only when the fully relaxed positions of a large number of neighbours are used, which forbids any analytical formulation of the solution and weighs the computational burden down by a appreciable amount.

**1.3.2. *Brute force method*.** The route is fully numerical like the supercell method; it rests entirely on the numerical knowledge of interatomic interactions as a function of distance. As a first step the fully relaxed positions are determined for all atoms in the cell; as a second step, the vibrational modes of the atoms belonging to the inner region are calculated, while keeping immobile the atoms of the outer region. It can be shown that the contribution of the outer region to the entropy must scale as 1/N* where N* is the number of atoms in the inner region [16]. Two approaches have been tried:

* a first procedure determines the limit of the entropy when N* is increased, the latter remaining bounded in order to keep the computational work to a tractable level [18];

* a second procedure keeps an inner zone of moderate extension but applies systematically a correction to account for the entropy change stemming from the lattice distortion of the outer zone. It is shown that the local pressure change on each site can be converted into a corresponding local expansion or contraction and further conveniently translated into an entropic contribution which improves noticeably the value of the total entropy [19].

If convenient for defective configurations of high symmetry like a vacancy, this last approach neglects the shear components of the distortion. These shears become no longer negligible for more complex defective configurations like a defect clusters or saddle points, the dipolar tensors of which exhibit clearly very different eigen values, whatever the type of potential (pairwise, many-body) used in the calculations [20-23]. As an illustration, the saddle configuration for a vacancy jump in a face-centered cubic lattice along <110> exhibits an orthorhombic symmetry, corresponding to a contraction of the lattice along <110> together with an extension along <001> ; depending on the detail of the potential, another contraction of smaller magnitude along <1-10> is exhibited [21].

The aim of the present paper is to go one step further along the line initiated in [19] and to propose a more refined method including the contribution of shear strains to vibrational entropies in the frame of the harmonic approximation. The paper is organized as follows: i) a first part recalls some basics about the link between strain, energy and entropy and defines the ingredients to be determined, namely the entropic constants; ii) a second part explains the way how the entropic constants are practically evaluated; iii) a third part illustrates this approach on copper, when modelled with several empirical potentials available in the litterature. Various analytic forms will be chosen with the hope of extracting generic features; we show in passing that the continuity of all the ingredients entering the formulation of the interactions and of their derivatives up to third order included is mandatory; iv) a fourth part explores briefly the problem raised by concentrated alloys. In order to make the paper not too obscure, only the main lines of the approach are described and presented in the body of the text, while rejecting all technical details in appendices.

## 1. Energy and entropy change under a uniform strain

### 1.1. *Energy change and elastic constants*

For a uniformly strained medium, the energy increase per unit volume, under zero external pressure and at 0 K, is given by:

$$\Delta U = \frac{1}{2} C_{ijkl} \eta_{ij} \eta_{kl} + \frac{1}{6} C_{ijklmn} \eta_{ij} \eta_{kl} \eta_{mn} + \frac{1}{24} C_{ijklmnop} \eta_{ij} \eta_{kl} \eta_{mn} \eta_{op} \cdots \quad (7)$$

where $C_{ijkl}$, $C_{ijklmn}$, $C_{ijklmnop}$ are the second, third, fourth ... order Brugger elastic constants [24] and $\eta_{ij}$ the components of the Lagrange strain tensor. Using hereafter Voigt's contraction of indices (ii => i ; 23 => 4; 31 => 5; 12 => 6) and limiting ourselves to cubic symmetry for sake of simplicity, the summation can be recast under a standard form displaying only the independent elastic constants, namely the three ones of second order, the six ones of third order and the eleven ones of fourth order [25], according to:

$$\begin{aligned}\Delta U = &\frac{1}{2}C_{11}P_{11} + C_{12}P_{12} + \frac{1}{2}C_{44}P_{44} \\ &+ \frac{1}{6}C_{111}P_{111} + \frac{1}{2}C_{112}P_{112} + C_{123}P_{123} + \frac{1}{2}C_{144}P_{144} \\ &+ \frac{1}{2}C_{155}P_{155} + C_{456}P_{456} + \frac{1}{24}C_{1111}P_{1111} + \frac{1}{6}C_{1112}P_{1112} \\ &+ \frac{1}{4}C_{1122}P_{1122} + \frac{1}{6}C_{1123}P_{1123} + \frac{1}{4}C_{1144}P_{1144} + \frac{1}{4}C_{1155}P_{1155} \\ &+ \frac{1}{2}C_{1244}P_{1244} + \frac{1}{2}C_{1266}P_{1266} + C_{1456}P_{1456} + \frac{1}{24}C_{4444}P_{4444} \\ &+ \frac{1}{4}C_{4455}P_{4455} + ....\end{aligned} \quad (8)$$

The $P_{ij}$, $P_{ijk}$, $P_{ijkl}$ are homogeneous polynomials of second, third and fourth order respectively of the Lagrange strains $\{\eta_i\}$. Their expressions can be found in [26] up to third order and in [27] for fourth order (apart from an error in some multiplicative factors ahead); they are recalled below in equations 9 for sake of completeness.

$$\begin{aligned}&P_{11} = \eta_1^2 + \eta_2^2 + \eta_3^2 \quad P_{12} = \eta_1\eta_2 + \eta_2\eta_3 + \eta_3\eta_1 \quad P_{44} = \eta_4^2 + \eta_5^2 + \eta_6^2 \\ &P_{111} = \eta_1^3 + \eta_2^3 + \eta_3^3 \\ &P_{112} = \eta_1\eta_2(\eta_1 + \eta_2) + \eta_2\eta_3(\eta_2 + \eta_3) + \eta_3\eta_1(\eta_3 + \eta_1) \\ &P_{123} = \eta_1\eta_2\eta_3 \qquad P_{144} = \eta_1\eta_4^2 + \eta_2\eta_5^2 + \eta_3\eta_6^2 \\ &P_{155} = \eta_1(\eta_5^2 + \eta_6^2) + \eta_2(\eta_6^2 + \eta_4^2) + \eta_3(\eta_4^2 + \eta_5^2) \quad P_{456} = \eta_4\eta_5\eta_6 \\ &P_{1111} = \eta_1^4 + \eta_2^4 + \eta_3^4 \\ &P_{1112} = \eta_1\eta_2(\eta_1^2 + \eta_2^2) + \eta_2\eta_3(\eta_2^2 + \eta_3^2) + \eta_3\eta_1(\eta_3^2 + \eta_1^2) \\ &P_{1122} = \eta_1^2\eta_2^2 + \eta_2^2\eta_3^2 + \eta_3^2\eta_1^2 \qquad P_{1123} = (\eta_1 + \eta_2 + \eta_3)\eta_1\eta_2\eta_3 \\ &P_{1144} = \eta_1^2\eta_4^2 + \eta_2^2\eta_5^2 + \eta_3^2\eta_6^2 \\ &P_{1155} = \eta_1^2(\eta_5^2 + \eta_6^2) + \eta_2^2(\eta_6^2 + \eta_4^2) + \eta_3^2(\eta_4^2 + \eta_5^2) \\ &P_{1244} = \eta_1\eta_2(\eta_4^2 + \eta_5^2) + \eta_2\eta_3(\eta_5^2 + \eta_6^2) + \eta_3\eta_1(\eta_6^2 + \eta_4^2) \\ &P_{1266} = \eta_1\eta_2\eta_6^2 + \eta_2\eta_3\eta_4^2 + \eta_3\eta_1\eta_5^2 \quad P_{1456} = (\eta_1 + \eta_2 + \eta_3)\eta_4\eta_5\eta_6 \\ &P_{4444} = \eta_4^4 + \eta_5^4 + \eta_6^4 \qquad P_{4455} = \eta_4^2\eta_5^2 + \eta_5^2\eta_6^2 + \eta_6^2\eta_4^2\end{aligned} \quad (9)$$

The elastic constants can be calculated to any order through lattice sums as soon as the interatomic potential is known, provided its analytical expression is everywhere continuous up to the necessary order (third order at least for $C_{ijk}$, fourth order at least for $C_{ijkl}$ ...etc) [28]. We report in Appendix A the formal expressions of the fourth and fifth order elastic constants as a function of the derivatives of the pair and N-body terms which will be used later on for the potentials under examination.

For sake of simplicity in its further use, the equation 8 is rewritten in a more compact form. Renaming the 20 elastic constants by $C_i$ ($C_{11} = C_2$, $C_{12} = C_3$, $C_{44} = C_4$ ...etc) and the 20 homogeneous polynomials by $P_i$ ($P_{11} = P_2$, $P_{12} = P_3$, $P_{44} = P_4$ ...etc) in the order of their introduction in equation 8 (including the fractional multiplicative term), the equation is transformed into:

$$\Delta U = \sum_{k=2}^{21} C_k P_k \tag{10}$$

### 1.2. *Entropy change and entropic constants*

The internal energy and entropy of a crystalline body must be represented by functional forms which are invariant through the symmetry operations of the lattice [29]; their expressions are functions of the same combinations of scalar invariants, the set of which is known for all lattices and symmetry groups [30]. As a consequence, the general expression for the entropy change under an homogeneous strain will look like the one for the energy, where the elastic constants $C_{ij}$, $C_{ijk}$, $C_{ijkl}$ will be replaced by 'entropic constants' $\lambda_{ij}$, $\lambda_{ijk}$, $\lambda_{ijkl}$. A difference must be noticed however. Concerning the energy, the reference state is most often taken as the equilibrium state under zero external pressure, the latter condition bringing no further contribution to the energy if the volume is increased during the application of the strain; therefore, the first order term which is proportionnal to the volume expansion is lacking in equation 8. Conversely, when considering the entropy, the dilatational part of the strain brings always a first order contribution proportionnal to the volume expansion [9]:

$$\partial S / \partial V \big|_P = \alpha B \tag{11}$$

where $\alpha$ is the thermal volume expansion coefficient, B the bulk modulus. Switching from derivatives to finite differences and denoting the volume of the system in the reference state by $V_o$, an alternate writing of this equation enlightens the link existing between the entropy change and any volume change in general, whatever its physical origin (temperature increase, elastic strain ...) :

$$\Delta S = \alpha B V_o \, \mathrm{dilat} = \lambda_1 \, \mathrm{dilat} \tag{12}$$

Let us recall here that in Lagrange's formalism, the volume expansion denoted here by 'dilat' is no longer expressed as the trace of the strain tensor but rather by:

$$\mathrm{dilat} = \left[ \mathrm{Det}(\mathbf{I} + 2\boldsymbol{\eta}) \right]^{1/2} - 1 \tag{13}$$

As above for the energy, the entropy change is written in a closed form. Renaming by $\lambda_k$ (k =1, 21) the entropic constants and by $P_1$ the volume expansion given by equation 13, the entropy change will later on be expressed as:

$$\Delta S = \sum_{k=1}^{21} \lambda_k P_k \tag{14}$$

### 2. Determination of entropic constants

The free energy of a system submitted to specific boundary conditions (external stress, temperature) can be determined in an elegant way by a direct minimisation of the total free energy yielding the equilibrium atomic configuration [31, 32]. The vibrational contribution to the quasiharmonic free energy is given as a function of an arbitrary wave vector **q** but a numerical summation over a relevant set of **q** vectors in the the first Brillouin zone is still necessary to obtain the final value of the entropy as a function of the strain amplitude. This is a convenient approach for determining the states of minimum free energy in systems with internal degrees of freedom. Since we are working in the harmonic approximation with a perfect system containing a single atom in its unit cell, a simpler method is sufficient: the homogeneous strain applied to the simulation cell involves no change of the inner atomic coordinates and the relaxation step becomes superfluous. This is the reason why we choose a more direct route: a set of predetermined strains of increasing amplitudes are imposed to the simulation cell and the entropy changes are numerically evaluated and collected; the desired entropic constants are then extracted through a least square fitting of the data to the theoretical behaviour depicted by equation 14.

## 2.1. *Definition of the set of strains*

Following Cousins [26] we define a stretching tensor **J** which is chosen rotation free by selecting a symmetrical form. The intensity of the components can be varied at will with respect to some arbitrary amplitude denoted hereafter by $\varepsilon$ :

$$\mathbf{J} = \begin{vmatrix} 1+d_1\,\varepsilon & d_6\,\varepsilon & d_5\,\varepsilon \\ d_6\,\varepsilon & 1+d_2\,\varepsilon & d_4\,\varepsilon \\ d_5\,\varepsilon & d_4\,\varepsilon & 1+d_3\,\varepsilon \end{vmatrix} \qquad (15)$$

The Lagrangian strain tensor $\boldsymbol{\eta}$ is then defined by $\boldsymbol{\eta} = \frac{1}{2}(\mathbf{J}^2 - \mathbf{I})$ where **I** stands for the unit (3×3) matrix. A volume expansion is involved whenever the determinant Det(**J**) is larger than unity and a contraction otherwise. As a consequence, for a given **J**, the strains defined by **J** (Det(**J**))$^{-1/3}$ or **J** will take place at constant volume or not, respectively. We gathered and numbered in table 1 the set of strains which were applied to the simulation cell. Some of the strains are pure shears inducing no volume change (n° 2, 4, 6, 9-26, 28, 30, 32, 34) ; several others imply a volume change of first order (n° 1, 3, 7, 29, 31, 33, 35) or second order (n° 5, 8, 27) with respect to $\varepsilon$. Several of them are redundant (11 and 12, 15 and 16 ...etc) and are expected to give similar results (they were subsequently checked that they do so). Many of these strains would be impossible to apply in a practical experiment, at variance with the simulation of a true elastic axial loading [33]: they must be viewed as thought experiments which are used as a simple and systematic way to probe numerically the entropy surface as a function of the strain variables $\{\eta_i\}$.

(insert table 1 around here)

The strains are applied to the three orthonormal basis vectors (**a, b, c**) defining the simulation box. As already mentionned above, the symmetry of the system implies that the reduced atomic coordinates need not to be changed; as a matter of fact, the conjugate gradient algorithm in charge of relaxing the atomic positions finds systematically that the system is already in its minimum energy. The product of the eigen modes is then determined in the deformed state, which involves only the evaluation of the determinant of the force constant matrix. The corresponding energy and entropy changes with respect to the perfect undeformed state are recorded.

The calculations are conducted in a cubic cell of FCC structure containing 864 atoms. To gather a significant amount of results, the amplitude $\varepsilon$ of the strain is varied in the range [-$10^{-2}$ : +$10^{-2}$] with 100 equally spaced mesh points between [± $10^{-2}$ : ± $10^{-4}$] and 100 between [± $10^{-4}$ : ±$10^{-6}$] for each type of strain.

## 2.2. Extraction of the entropic constants

After the energy and entropy changes have been collected for all types of strains and all amplitudes, the elastic and entropic constants were determined by a least square fitting of equations 10 and 14.

For strain amplitudes amounting up to a fraction of a percent, it could be checked that the fourth order terms in equation 10 are necessary. If equation 10 is reduced to the second order terms ($\lambda_{i\leq 4}$), energy changes are reproduced with an average relative precision of only a few $10^{-2}$; if third order terms are used ($\lambda_{i\leq 10}$), the relative precision is better than $10^{-3}$; if fourth order terms are used, the precision improves further down to $10^{-4}$. We checked further that including fith order terms did not improve the fit in a detectable way. The fourth order formula was therefore considered as the most convenient and used systematically.

The fitted elastic constants were systematically compared to those obtained at equilibrium through lattice sums [28]. The precision of the agreement was found to depend on the potential and on the order of the elastic constant. The potentials which were tried differ mainly by their interaction ranges and will be examined in more detail below but general features can be extracted now. The agreement between the fitted and the equilibrium values of the elastic constants followed the following trend: i) for second order elastic constants, the relative precision was better than $10^{-4}$ for all potentials; ii) for third order elastic constants, the relative precision ranged from $10^{-3}$ for the longer-ranged potential to a few percents for some others; iii) for fourth order elastic constants, the relative precision was a few percents for the longer-ranged potential ; for some others only the signs and the order of magnitude were correct. As a general trend, the longer the range of the interaction, the better the agreement for the higher orders. The fact that all elastic constants were reasonably reproduced was interpreted as an indirect proof that the types of strains included in the set was broad enough to probe correctly the potential energy surface of deformed states.

The corresponding entropic constants were determined in the same way. As a general rule, the entropic changes were reproduced through equation 14 with a relative accuracy of the same order as that for the energy. The entropic constants exhibited however a higher sensitivity to the fitting procedure than the elastic ones. For instance $\lambda_1$ and $\lambda_4$ are rather insensitive to the maximum order retained in equation 14 or to the strain amplitudes used in the fitting, whereas $\lambda_2$ and $\lambda_3$ could vary of 20% . This is the reason why we consider the entropic constants as defined to no better than $10^{-4}$ for the first order constant $\lambda_1$, a few $10^{-2}$ for the second order constants $\lambda_2 = \lambda_{11}$, $\lambda_3 = \lambda_{12}$ and $\lambda_4 = \lambda_{44}$ and a few $10^{-1}$ for the third order ones $\lambda_5$ to $\lambda_{10}$. The fourth order ones could not be determined safely.

## 2.3. Use of the entropic constants in defective configurations

The coefficients in equation 14 were determined for homogeneously strained systems. We assume that the same type of formulation can be applied locally, that is, on each of the N-N* sites of the outer region where the values of the polynomials $P_k$ will now depend on the local strain measured on site 'i'. The total entropy change $\Delta S_{out}$ of the outer region will thus include a summation running over the N-N* lattice sites:

$$\Delta S_{out} = \sum_{i=N^*+1}^{N} \sum_{k=1}^{21} \lambda_k \left[ P_k \right]_i \tag{16}$$

The next point to solve is to find a consistent and stable procedure for determining the local strain. At the end of a relaxation calculation, the energy of the defective system is known with a precision of the order of $10^{-12}$ at best and the atomic coordinates are not known to better than $10^{-6}$. The local strains must then be larger than a few $10^{-6}$ to be detected with some confidence. Using the atomic coordinates to calculate the local strain is not convenient: the discrete and anisotropic nature of the lattice at short scale complicates the calculation of the distortions and numerical derivatives of the various displacements must be taken, leading to an important numerical noise. We propose alternatively to evaluate the local stresses, the calculation of which has become a standard in atomistic calculations [34] and to extract the strains from the knowledge of the stresses. The calculated strain in this procedure becomes much less sensitive to the numerical noise on individual atomic coordinates since it rests on some average performed over all the neighbours interacting with a given atom. We recall below the classical way of determining numerically the stress components and the formal expressions of stress versus strain components.

We will deal later only with interactions describing the total energy of the atomic assembly as well as the energy per atom 'i' by:

$$U = \frac{1}{2} \sum_i \sum_{i \neq j} V(r_{ij}) - \sum_i F(\rho_i) = \sum_i U_i \tag{17}$$

where $V(r_{ij})$ stands for the pairwise part, F for the embedding function and $\rho_i$ is a local density using a pairwise function $\Phi(r_{ij})$ and given by :

$$\rho_i = \sum_{i \neq j} \Phi(r_{ij}) . \tag{18}$$

The component of the local stress on site 'i' is given by:

$$\sigma^i_{\alpha\beta} = -\frac{1}{\Omega_i} \frac{\partial U_i}{\partial \eta_{\alpha\beta}} = -\frac{1}{2\Omega_i} \sum_{j \neq i} \left( \frac{\partial V}{\partial r_{ij}} - \frac{\partial F}{\partial \rho_i} \frac{\partial \Phi}{\partial r_{ij}} \right) \frac{r^\alpha_{ij} r^\beta_{ij}}{r_{ij}} \tag{19}$$

where $\Omega_i$ stands for the atomic volume attached to site 'i' and $r^\alpha_{ij}$ for the '$\alpha$' component of vector $r_{ij}$. All these quantities can be numerically determined for each site 'i' and the stress components have been derived assuming that $\Omega_i$ is equal to the average atomic volume $\Omega_o$.

The theoretical expression of the stress component is readily obtained by deriving equation 8 with respect to the relevant strain component. Expressing '$\alpha\beta$' with Voigt's index 'I' and using equation 10 yields :

$$\sigma^i_I = \sum_{k=2}^{21} C_k \left[ \frac{\partial P_k}{\partial \eta_I} \right]_i \tag{20}$$

The presence of higher order terms forbids a linear inversion of equation 20. The components of the local strain are thus adjusted by a conjugate gradient procedure minimising the sum of the squared departures between equation 19 and 20. The solution of the linearised problem, resting on the use of only $C_{11}$, $C_{12}$ and $C_{44}$, is taken as the starting point of the iterative process.

It is worth noticing here that the approach described in [19] assumes a linear relationship between stress and strain; this assumption allows to convert directly the local pressure change $\Delta p_i$ into a local volume change, which is translated afterwards into a local entropic contribution through equation 12:

$$\Delta S_{out} = -\frac{\lambda_1}{B} \sum_{i=N^*+1}^{N} \Delta p_i , \qquad (21)$$

where the summation runs on the N-N* atoms of the outer region. It is worth mentionning that equation 21 can be written in an alternate form. Since our relaxations are performed under zero external pressure, it means that the average internal pressure calculated from a summation over the internal stresses is also zero. As a consequence the summation above can be replaced by a summation with an opposite sign running only on the atoms of the inner region:

$$\Delta S_{out} = +\frac{\lambda_1}{B} \sum_{i=1}^{N^*} \Delta p_i , \qquad (22)$$

which is faster since, in the spirit of the approximation, N* is supposed to be markedly smaller than N. And we checked extensively all along our calculations that the two results were indeed identical. As already noticed above, this approximation ignores the fact that a pure shear at constant volume in a solid (unlike a liquid) gives rise to a pressure change which varies as the square of the strain and brings a contribution of the same order as the one brought by the shear components. The pressure change induced by the shears is thus unduely attributed to a local expansion or contraction. However the two entropic contributions do not behave similarly: as will be seen later, the contribution of shears is most often positive, whereas that of pressure has both signs. It is thus very important to know which one brings what.

We will apply our approach to the formation of single-di-and-tri vacancy and of dumbbell interstitial as well as to the migration jump of these defects. The entropy difference between the perfect state and the defective one is made of two contributions. The contribution $\Delta S_{in}$ of the inner region is given by:

$$\Delta S_{in} \simeq -\frac{k}{2} \text{Log}(\Delta'/\Delta^{(N^*-N_d)/N^*}) \qquad (23)$$

where N* and N*-$N_d$ denote the number of atoms in the inner region of the perfect and defective state respectively ($N_d$ = -1, -2, -3 for a single-di-tri vacancy and +1 for a dumbbell). The contribution $\Delta S_{out}$ of the outer zone is given by equation 14.

The pre-exponential factors $\nu_o$ of the jump frequencies are evaluated according to transition rate theory [35-36] for the inner region, the eigen modes of which must be calculated in the starting stable configuration and in the saddle configuration at the top of the energy barrier. The elastic correction is then applied to each site of the outer region after calculating the corresponding entropy change when going from the stable to the saddle configuration:

$$\nu_o = \frac{1}{2\pi} \frac{\prod_{i=1}^{3(N^*-N_d)} \omega_i^{stable}}{\prod_{i=1}^{3(N^*-N_d)-1} \omega_i^{saddle}} \exp\{(\Delta S_{out}^{saddle} - \Delta S_{out}^{stable})/k\} \qquad (24)$$

For the case of multi defects, the only jump studied here is non dissociative: the direction and the jumping partner are chosen in such a way that the multidefect structure is restored (apart from a rotation) after the jump has been completed.

In all cases, the atomic configurations used for the calculation of entropies are the results of a static relaxation involving all the atoms of the simulation cell (N=2048 atoms); the relaxation is conducted under a constant external zero pressure as described in [3]. In the final stable or saddle states, the component of the forces are of the order of $10^{-6}$ eV Angs$^{-1}$. The size of the inner region ranges from 55 to 791. The entropies are also calculated in the supercell approach for N = 256, 500, 864, 1372 and 2048.

## 3. Introduction of the empirical potentials used in the study and results

Before reviewing the four potentials used in this work, we draw the reader's attention on an important preliminary point which has been already mentionned by previous authors, namely the continuity of interactions at higher order [e.g. 15]. It is shown below that a discontinuity of third order can produce systematically faulty values of entropies. We illustrate this point with a simple potential of Finnis-Sinclair type in the subsequent section.

### *3.1. Possible deleterious effects of a third order discontinuity on the entropy*

The energy is expressed by equations 17 and 18 with $V(r_{ij}) = \alpha \exp(-p\, r_{ij}/r_o)$, $\Phi(r_{ij}) = \exp(-2q\, r_{ij}/r_o)$ and $F(\rho_i) = \beta \rho_i^{1/2}$. $r_o$ stands for the first neighbour distance in the perfect crystal at rest. The four adjustable parameters $\alpha$, $p$, $\beta$, $q$ are fitted on the cohesive energy $E_{coh}$, the bulk modulus B, the lattice constant $a_o$ at rest (zero temperature and pressure), the unrelaxed formation energy of the vacancy $E_f$, while taking into account the first, second and third neighbour shells only. Truncating the potential between third and fourth neighbours consists in replacing the exponential decay by a polynomial of fifth order, between some distance $r_{rac}$ arbitrarily chosen beyond the third neighbour distance ($r_{rac} = 1.825\, r_o$) and the cutoff distance ($r_{cut} = 2\, r_o$) of the form

$$\text{Pol}(x) = ax^3 + bx^4 + cx^5 \qquad (25)$$

where x stands for r-$r_{cut}$. The polynomials used for V(r) and $\Phi$(r) will be called P(x) and P'(x) respectively: the coefficients a, b, c for P(x) and a', b', c' for P'(x) are determined in such a way that the derivatives are continuous up to second order included at $r_{rac}$. This potential fitted on copper ($E_f$ =1.28 eV, $E_{coh}$=3.54 eV, $a_o$=3.615 Angs, B=0.864583 eVAngs$^{-3}$) is called FS-PT345-2.00 ('PT' for 'polynomial termination'; '345' for the exponents; 2.00 for the value of $r_{cut} / r_o$). The numerical values of the coefficients are gathered below in table 2.

(insert table 2 around here)

A simulation cell containing 864 atoms is then strained by a pure shear involving no volume change (strain n°13 in table 1 with an amplitude $\varepsilon \leq 10^{-3}$) and the calculated entropy change in units of k is plotted versus the square of the shear strain amplitude $\varepsilon^2$ in figure 1. The expected quadratic behaviour is not obtained and a quasi linear variation is observed (top curve of figure 1 with a slope equal to 1/2). It is however easy to check that a new calculation, either with $r_{cut}$ set to 1.99 $r_o$ (FS-PT345-1.99), or after replacing Pol(x) by a polynomial of higher order, namely $ax^4 + bx^5 + cx^6$ (FS-PT456-2.00 and FS-PT456-1.99) cures entirely the anomaly and restores for the three lower curves in figure 1 the expected slope. The curves corresponding to the same cutoff radius 1.99 are superposed because the difference of the terminating polynomials does not play any role; the curve corresponding to the cutoff 2.00 is slightly higher since it establishes a vibrational connection between more distant neighbours.

Similar conclusions are reached for the formation entropies of a vacancy or a dumbbell: the potential FS-PT345-2.00 exhibits anomalously large values. The anomaly is cured by using the other versions of the potential. It is worth noticing that, in all cases, the energies are very close to one another (table 3).

(insert table 3 around here)

This faulty behaviour comes from the fact that the vibrational entropy is a function of the second derivatives of the interactions, and that a distortion of the lattice lets implicitly come into play third order derivatives. In the present case, the origin of the flaw lies in the polynomial termination through which the interactions decay to zero: this polynomial has a discontinuity of its third derivative at $r_{cut}$ which was set equal to the fourth neighbour distance exactly. For the homogeneous shear strain under examination, some of the fourth neighbours enter or leave the interaction range: these displacements include unfortunately the discontinuity and induce an abrupt variation of the force constants. The flaw is cured, either by choosing a slightly smaller cutoff radius keeping the fourth neighbours out of reach, or by using a polynomial of higher order which changes the discontinuity at $r_{cut}$ from third to fourth order. Let us notice that we did not suppress the discontinuity of the third order derivative around $r_{rac}$, but the latter distance is too far from any of the neighbour distances involved in the calculation of perfect or defective configurations to spoil the results.

It could obviously be argued that the choice of the cutoff distance in FS-PT345-2.00 is very clumsy. Indeed, around single point defects, the strains of the embedding lattice are known with a good precision and the possible discontinuities of the potential (cutoff distance for the physical interactions, radius where an analytical form switches to another ...etc) can be carefully tuned in such a way that none of the neighbour shells will ever cross these points during the lattice deformation leading from the perfect reference state to the defective one. The choice is much more difficult, or even impossible, when dealing with situations involving large strains around complex defects like those produced by saddle configurations, interstitial or vacancy clusters, coherent interfaces with a large size misfit, grain-boundaries or local rearrangements during the collapse of a vacancy cluster into a vacancy loop ...etc. In all these cases, one can never be sure that some shell of neighbours will not cross a point where a third derivative of the potential is discontinuous. This is the general reason why analytical formulations of interactions should be preferentially chosen to avoid any discontinuity of any order.

It is worth noticing that this basic requirement of continuity was not always met in practice in previous investigations: although the basic interactions are evaluated with more and more sophisticated data including ab initio results, the quality of the final representation can be damaged by a functional expression which does not meet the requirement of continuity for high order derivatives. For instance, in [37] the embedding function is represented by cubic splines, the latter introducing as many discontinuities of the third order derivative as the number of mesh points; in [38] the embedding function is represented by two different expressions on the two sides of the equilibrium density $\rho_0$ with a discontinuity of the third order derivative at $\rho_0$; in [39] the embedding function stems from an equation of state represented by two different expressions on the two sides of the equilibrium lattice parameter $a_0$; in this last case fortunately, the third order terms are identical on the two sides and the discontinuity shows up only at the fourth order derivative.

### 3.2. *Empirical potentials and related results*

The reason why we used several empirical potentials is not for the purpose of a quantitative comparison, which would have a minor interest: everybody knows that all of thems are basically wrong since none rests on a formulation of cohesion and atomic forces taking full account of the electronic structure. Their only virtue is to propose a simple phenomenological formulation of the interaction which allows a reasonable fit to some set of selected physical properties and experimental results. All the potentials examined below obey the general form of equations 17 and 18 above, but their pairwise and density function terms are described by very different analytical expressions and are fitted to different physical properties.

Two potentials are of Finnis Sinclair type: i) the first one has a long interaction range with an exponential damping function, is fitted only to four physical properties of bulk copper quoted in section 3.1 and called hereafter FS-ED-2.82; ii) the second has a very short interaction range, is fitted to vibrational properties of copper surfaces [40] and called hereafter FS-Barreteau.

Two potentials are of EAM type: i) the first one is fitted to various properties of bulk copper including ab initio data [38]; we explain in Appendix B the way how it was modified to meet the requirement of continuity for its third derivative. This modified potential is called EAM-Mishin-mod; ii) the second one is explicitely fitted to elastic constants of second and third order [39] which have been experimentally determined for long [41] and called EAM-Milstein.
All the numerical details related to each potential are reported in Appendix B, including the resulting values of the elastic constants up to fourth order, the energetic and entropic parameters attached to the point defects under examination.

### 3.3. *Determination of entropic constants for the four potentials*

The entropic constants were determined using the procedure described in section 2.2. and are gathered in table 4:

* the first order entropic constant $\lambda_1$ has a comparable magnitude for the four potentials: it means that similar expansion thermal expansion coefficients are predicted;

* if $\lambda_2$ and $\lambda_3$ are different from one potential to the other, the difference $\lambda_2$-$\lambda_3$ has the same sign and the same order of magnitude, at least for the first three of them; EAM-Milstein is a special case and gives a larger result by one order of magnitude;

* unlike $\lambda_2$ and $\lambda_3$, the second order entropic constant $\lambda_4$ has a comparable order of magnitude, except for the potential EAM-Mishin-mod which yields a value twice smaller;

* the entropic constants $\lambda_{i>4}$ corresponding to higher order terms are very different in magnitude and sign and do not exhibit any noticeable trend; fortunately, as will be verified below, they bring a hardly detectable contribution to the entropy change.

Although the potential EAM-Milstein was more carefully fitted to elastic properties than the others, it produces formation entropies of vacancy defects which are roughly twice larger than those evaluated by the other potentials, whereas the formation entropy for dumbbell formation is similar. No explanation was found for this behaviour.

(insert table 4 around here)

A last point is worth mentioning: for all the potentials but one, pure shear strains at constant volume produce positive entropy changes only. The exception is EAM-Mishin-mod which exhibits positive entropy changes for only four strains (n° 2, 4, 6, 9) and negative ones for all others (n° 10-26, 28, 30, 32, 34). Brillouin has demonstrated that the equality of isothermal and adiabatic shear moduli G implied a quadratic dependence of the entropy as a function of the shear amplitude $\varepsilon$ [29] of the form $S \simeq m \varepsilon^2$ where $m = -dG/dT$ is expected from physical arguments to be positive. This conclusion is however established for an isotropic solid with linear elasticity. For copper, two shear moduli can be defined, namely $(C_{11}-C_{12})/2$ and $C_{44}$, which differ from each other by a factor of 3. It is however readily checked for this potential that their temperature derivatives are negative as they should be; but higher order terms might be important. We discuss in Appendix C a possible ingredient of the interatomic interactions which might influence the sign of the entropy change.

### 3.4. *Effects of the elastic correction including shear strains*

We pass in review the effects of our elastic corrections on the defective configurations enumerated in section 2.3. We compare the entropies obtained by the embedded crystallite method in its original form (EC), with the elastic correction including pressure (ECp) and with our elastic correction including dilatation and shears (ECds). All the results are rated against those given by the supercell method (SC). The results obtained with the potentials FS-ED-2.82, FS-Barreteau, EAM-Mishin-mod and EAM-Milstein will be displayed systematically in this order.

The form of equation 11 and 14 dictates the general trend to be expected: the magnitude and sign of the global entropic correction depends mainly on the magnitude and on the sign of the relaxation volume between the reference state and the defective one, since it provides the first order term. The additional effect of shear strains is of second order and shows up only in those configurations where shear stresses (and strains) are present with a noticeable magnitude.

A last remark should be made here about the way with which the calculations are used, as well as about the relative importance of the ingredients entering the final results. The sites on which the elastic correction is applied are never in the close vicinity of the defect producing the strain; as a consequence, smaller strains than those used for determining the entropic constants are experienced, ranging typically from a few $10^{-3}$ to a few $10^{-6}$. However, we checked on preliminary calculations conducted on saddle configurations that the local strain tensor $\eta$ could not be extracted with an assumption of linear relationship between stress and strain. In other words the expression of the energy taking into account only the second order terms corresponding to the first line in equation 8 is not sufficient and higher order terms are mandatory. As an illustration, we compare below in table 5 the entropic corrections evaluated in the stable and saddle configurations obtained with the potential FS-ED-2.82: i) with the only knowledge of the local pressure as proposed by equation 21 (or 22) and denoted by $\Delta S_{out}^{stable}(press)$ and $\Delta S_{out}^{saddle}(press)$; ii) with the dilatation contributions $\Delta S_{out}^{stable}(dilat)$, $\Delta S_{out}^{saddle}(dilat)$ and shear contributions $\Delta S_{out}^{stable}(shear)$, $\Delta S_{out}^{saddle}(shear)$ as proposed by equation 14 of our approach.

For the latter, a preliminary remark should be made. The total entropy increase defined by equation 14 can be unambiguously partitionned into a first contribution $\Delta S_{O1}$ coming from the unique first order term and a second contribution $\Delta S_{O>1}$ grouping higher order terms all together. Conversely its decomposition into contributions stemming from dilatation and shears respectively is somewhat arbitrary for two reasons: i) working with non linear elasticity, the definition of the pure dilatation part $\{\eta_i\}^d$ of the initial strain $\{\eta_i\}$ according to $\{\eta_i\}=\{\eta_i\}^d + \{\eta_i\}^s$, defines a second term $\{\eta_i\}^s$ which is no longer a pure shear strain because it contains dilatation terms of second order or more; ii) the fact that equation 14 contains non linear terms prevents us from decomposing formally the total entropy increase into two independent contributions corresponding to $\{\eta_i\}^d$ and $\{\eta_i\}^s$ respectively, because cross terms will unavoidably show up which mix components of $\{\eta_i\}^d$ and $\{\eta_i\}^s$. In our case (small strains around point defects), we can however verify that these cross terms have a negligible contribution as follows. With equation 14, we can calculate independently from each other the entropic contributions $\Delta S(\text{dilat})|_{\eta^d}$ and $\Delta S(\text{shear})|_{\eta^s}$ due to a pure dilatation $\{\eta_i\}^d$ and an impure shear $\{\eta_i\}^s$ respectively. It is then checked readily that $\Delta S(\text{dilat})|_{\eta^d}$ and $\Delta S(\text{shear})|_{\eta^s}$ are nearly equal to $\Delta S_{O1}$ and $\Delta S_{O>1}$ respectively, to better than one percent. This is the reason why the first order term in equation 14 will be considered, at least quantitatively, as the contribution of dilatation and displayed as such in subsequent table 5.

The dilatational and shear contributions are calculated with stresses derived from an expression of the energy taking into account second order terms, second and third order terms and second, third and fourth order terms in equation 8, the corresponding results being denoted by O(2), O(3) and O(4) respectively. The expressions used for the entropic correction are of the corresponding order.

(insert table 5 around here)

For the stable configuration, the final correction is negative, as expected, since the relaxation volume for vacancy formation is negative. The results O(2), O(3) and O(4) are very close. The ECp and ECds approaches give the same result, which was a priori not obvious. In a continuous medium treated with isotropic elasticity the shear strains around a spherical hole have only tangential components. In the discrete medium made up of a FCC lattice, the pattern of internal stresses is more complicated: the dense rows <110> running through the vacancy site are under tension while the others are compressed, thus leading to non zero radial shear components. But their influence on the final result is apparently very weak.

For the saddle configuration, a detailed inspection of the atomic sites shows that extracting the strains from stresses through formulas of increasing order enhances (algebraically) the local dilatation and decreases the weight of its negative contribution compared to that of shears. The final contribution due to dilatation remains however negative: indeed, the lattice parameter for the saddle point configuration is smaller than the parameter for the perfect lattice because the positive migration volume does not compensate entirely the negative relaxation volume observed during the vacancy formation. The contribution of the shears is conversely rather insensitive to the order of the approximation, which is probably due to the small strains experienced locally. As a result, the final entropic correction ECds is found negative with O(2) but it becomes positive with O(3); the additional contribution brought by O(4) terms is not detectable. This correction ECds is to be compared to the ECp one, which is found negative. It illustrates the point already mentionned above in section 2.3: in [19], the change of pressure is assumed to stem only from a dilatation and it neglects the fact that shears bring also a quadratic contribution to pressure. In the particular case of a saddle point, the contribution of the shears becomes even larger than the contribution of dilatation.
All our results displayed below and denoted by ECds are obtained after extracting the deformation through O(4) and evaluating the entropies with $\lambda_{i \leq 4}$.

**3.4.1. *Formation entropies.*** The entropic corrections defined above are displayed in figure 2 and 3 for mono- and tri-vacancy and figure 4 for the dumbbell interstitial.

For vacancy defects, the EC values are higher than the SC values. The relaxation volume is negative: as a consequence, the first order term of the elastic correction due to pressure or dilatation will bring a negative contribution. This correction makes the ECp and ECds results definitely closer to those obtained by supercell method, with some overshoot. For the single vacancy, the effect of shears is negligible which means that the vacancy behaves elastically as a contraction center of spherical symmetry: the ECds results are practically superposed upon the ECp ones (figure 2). The lower symmetry of di- and tri-vacancy gives rise to a small but detectable contribution of shears which gives a further positive correction and tends to decrease the overshoot by a fraction of k (figure 3 for the tri-vacancy).

For the dumbbell, the effect is reversed and the EC values are much smaller than the SC ones. The first order term of the elastic correction stemming from the positive relaxation volume is positive and very large: the ECp values are higher than the EC ones by several k (figure 4). The ECds values are slightly higher than the ECp ones by one k at most. The fact that the effect of shears is not very important is not too surprising and could have been predicted: indeed, although dissociated along a well defined direction, the dumbbell defect produces noticeable shear strains only in its close vicinity; the distortion of the lattice at larger distances is very close to what would be produced by a center of dilatation of sperical symmetry, a feature which is reflected by three nearly equal eigen values of its elastic dipole tensor [20].

It is worth noticing that the amplitude of the elastic corrections are comparable for the four potentials, although the absolute values of the entropies (with or without the corrections) are noticeably different from one another.
In all cases, as expected, the second order correction due to shears decays more rapidly than the first order correction due to dilatation and is visible only for the smallest sizes of the inner region: it becomes hardly detectable as soon as N* becomes larger than 500.
(insert figures 2,3,4 around here)

**3.4.2. *Pre-exponential factor of jump frequencies.*** Saddle configurations are good candidates for involving large shear strains; but different situations are met according to the nature of the jumping defect. Since the different potentials obtained for all defects qualitatively similar results, we show the migration results on figure 5 for one of them only (EAM-Mishin-mod).

For the single vacancy, the migration volume is positive, and the relative position of the data points on the graph is the same as the one observed for the dumbbell formation. But now the additional contribution of the shears, for the smallest sizes of the inner region, is of the same order of magnitude as the contribution of pressure or dilatation alone.

For the dumbbell, the migration volume is positive but very small. The same qualitative behaviour as the vacancy case is obtained but the overall change of the pre-exponential factor being less than 40% over the whole range swept by 1/N*, the graph does not look spectacular (figure 5b).

For the di-vacancy and the tri-vacancy, the migration volumes are negative and the graphs look like those displayed for the formation of these defects (figure 5c and 5d). The effect of shears is more pronounced for the divacancy than for the tri-vacancy. As noticed in Appendix B for the saddle point configuration of the tri-vacancy jump, the jumping atom is close to the center of the tetrahedron made up by the three vacancies and the atom on their starting stable sites: this restores a more spherical symmetry for the radial displacements and decreases correspondingly the shears.

(insert figure 5 around here)

For all the cases investigated here we can conclude that the effect of shears can be important in configurations of low symmetry. This conclusion is weakly dependent on the interatomic potential.

## 5. Conclusion

The entropy change under a general strain can be expressed as a function of the strain components at all order with the help of entropic constants, which play for the entropy the same role as the elastic constants for the energy.

The use of non linear elasticity implies that a pressure change can be due to a dilatation or to a shear strain: it destroys the one-to-one correspondence between pressure and dilatation on one hand and shear stresses and shear strains on the other, which was the rule for linear elasticity. As a consequence the canonical variables to be chosen for expressing energy and entropy changes are the strain rather than the stress components.

Using the approximation of the embedded crystallite for the calculation of vibrational entropies attached to defective configurations, it has been confirmed that the elastic correction associated with the distorsion of the surrounding matrix is of noticeable importance.

Although the first order dilatational term brings in all cases a leading contribution, higher order term due to shear strains adds a noticeable improvement in all the situations of lowered symmetry, namely all those distorted configurations around defects or defect clusters in their stable or saddle configuration. The contribution of shear strains can be of the same order of magnitude as the dilatation one for saddle configurations.

These results are obtained with several empirical potentials, which differ from one another by various analytical formulations of their ingredients or by their interaction range. It suggests that the effect we enlightened in this study is general.

## Aknowledgements

We thank E. Clouet and C. Marinica for valuable discussions and suggestions and J.P. Poirier (Académie des Sciences) for drawing our attention on Brillouin's papers.

# Appendix A: Calculation of 4<sup>th</sup> and 5<sup>th</sup> order elastic constants through lattice summations

For the general form of the potential adopted here, the energy per atom 'i' is assumed to be expressed as:

$$U_i = \frac{1}{2}\sum_{i \neq j} V(r_{ij}) - F(\rho_i) \tag{A1}$$

where $F(\rho_i)$ is some (non linear) embedding function of the local density $\rho_i$, the latter being a sum over interacting neighbors $\rho_i = \sum_{i \neq j} \Phi(r_{ij})$. $V(r)$ and $\Phi(r)$ stand for two-body interactions.

The n<sup>th</sup> order elastic constants are defined as the n<sup>th</sup> derivatives of the energy with respect to the components of the applied strain and can be written synthetically under the form:

$$C_{\underbrace{IJK...}_{n}} = \frac{1}{\Omega_O}\frac{\partial^n U}{\partial \eta_I \partial \eta_J \partial \eta_K ...} \tag{A2}$$

where $\Omega_O$ stands for the atomic volume in the reference state and I, J, K are the Voigt indices. The derivatives $\frac{\partial}{\partial \eta_I}$ are calculated via space derivatives $\frac{\partial}{\partial r}$ of the functions V and $\Phi$, where r stands for the distance between some lattice site chosen as the origin and any other lattice site. If $\vec{r}_O = x_1\vec{a} + x_2\vec{b} + x_3\vec{c}$, where $(\vec{a},\vec{b},\vec{c})$ stands for some orthonormal set of basis vectors, is changed through the strain into $\vec{r} = x_1\vec{a}' + x_2\vec{b}' + x_3\vec{c}'$, then the change of distance can be expressed in a condensed form as $r^2 - r_O^2 = 2\sum_I X_I \eta_I$ where $X_I$ stands for the product of coordinates $x_k x_l$. From this expression stems the derivative $dr/d\eta_I = X_I/r$.

The calculation of the first derivatives of the energy U, which are nothing but the internal stresses resulting from the applied strain, is then straightforward:

$$\Omega_O \sigma_I = \frac{\partial U}{\partial \eta_I} = \frac{1}{2}\sum_r \frac{dV}{dr}\frac{dr}{d\eta_I} - \frac{dF}{d\rho}\frac{d\rho}{d\eta_I} = \frac{1}{2}\sum_r \frac{dV}{dr}\frac{dr}{d\eta_I} - \frac{dF}{d\rho}\sum_r \frac{d\Phi}{dr}\frac{dr}{d\eta_I} \tag{A3}$$

where the summations over 'r' are to be performed on all neighbours which interact with the central atom chosen as the origin. Hence the final formal expressions:

$$\Omega_O \sigma_I = \frac{1}{2}SV_I - \frac{dF}{d\rho}S\Phi_I \tag{A4}$$

with the condensed notations

$$SV_I = \sum_r \frac{dV}{dr}\frac{dr}{d\eta_I} \qquad S\Phi_I = \sum_r \frac{d\Phi}{dr}\frac{dr}{d\eta_I}. \tag{A5}$$

Denoting the products $X_I X_J$ by $X_{IJ}$, $X_I X_J X_K$ by $X_{IJK}$ ... and extending the condensed notation to higher order derivatives of V and $\Phi$ according to :

$$Sf_{IJ} = \sum_r \left(\frac{d^2f}{dr^2} - \frac{1}{r}\frac{df}{dr}\right)\frac{X_{IJ}}{r^2}$$

$$Sf_{IJK} = \sum_r \left(\frac{d^3f}{dr^3} - \frac{3}{r}\frac{d^2f}{dr^2} + \frac{3}{r^2}\frac{df}{dr}\right)\frac{X_{IJK}}{r^3}$$

$$Sf_{IJKL} = \sum_r \left(\frac{d^4f}{dr^4} - \frac{6}{r}\frac{d^3f}{dr^3} + \frac{15}{r^2}\frac{d^2f}{dr^2} - \frac{15}{r^3}\frac{df}{dr}\right)\frac{X_{IJKL}}{r^4}$$

$$Sf_{IJKLM} = \sum_r \left(\frac{d^5f}{dr^5} - \frac{20}{r}\frac{d^4f}{dr^4} + \frac{45}{r^2}\frac{d^3f}{dr^3} - \frac{105}{r^3}\frac{d^2f}{dr^2} + \frac{105}{r^4}\frac{df}{dr}\right)\frac{X_{IJKLM}}{r^5}$$

(A6)

where 'f' stands for the V or $\Phi$ function. The expressions of the second, third, fourth and fifth order elastic constants are then easily deduced:

$$\Omega_O C_{IJ} = \frac{1}{2}SV_{IJ} - \frac{dF}{d\rho}S\Phi_{IJ} - \frac{d^2F}{d\rho^2}S\Phi_I S\Phi_J \qquad (A7)$$

$$\Omega_O C_{IJK} = \frac{1}{2}SV_{IJK} - \frac{dF}{d\rho}S\Phi_{IJK}$$
$$- \frac{d^2F}{d\rho^2}\left[S\Phi_I S\Phi_{JK} + S\Phi_J S\Phi_{KI} + S\Phi_K S\Phi_{IJ}\right] \qquad (A8)$$
$$- \frac{d^3F}{d\rho^3}S\Phi_I S\Phi_J S\Phi_K$$

$$\Omega_O C_{IJKL} = \frac{1}{2}SV_{IJKL} - \frac{dF}{d\rho}S\Phi_{IJKL}$$
$$- \frac{d^2F}{d\rho^2}\begin{bmatrix} S\Phi_I S\Phi_{JKL} + S\Phi_J S\Phi_{KLI} + S\Phi_K S\Phi_{LIJ} + S\Phi_L S\Phi_{IJK} \\ S\Phi_{IJ}S\Phi_{KL} + S\Phi_{JK}S\Phi_{LI} + S\Phi_{KI}S\Phi_{JL} \end{bmatrix}$$
$$- \frac{d^3F}{d\rho^3}\begin{bmatrix} S\Phi_{IJ}S\Phi_K S\Phi_L + S\Phi_{IK}S\Phi_J S\Phi_L + S\Phi_{IL}S\Phi_J S\Phi_K + \\ S\Phi_{JK}S\Phi_I S\Phi_L + S\Phi_{JL}S\Phi_I S\Phi_K + S\Phi_{KL}S\Phi_I S\Phi_J \end{bmatrix} \qquad (A9)$$
$$- \frac{d^4F}{d\rho^4}S\Phi_I S\Phi_J S\Phi_K S\Phi_L$$

$$\Omega_O C_{IJKLM} = \tfrac{1}{2} SV_{IJKLM} - \frac{dF}{d\rho} S\Phi_{IJKLM}$$

$$-\frac{d^2F}{d\rho^2}\begin{bmatrix} S\Phi_{IJK}S\Phi_{LM} + S\Phi_{IJL}S\Phi_{KM} + S\Phi_{IJM}S\Phi_{KL} + S\Phi_{IKL}S\Phi_{JM} + S\Phi_{IKM}S\Phi_{JL} \\ +S\Phi_{ILM}S\Phi_{JK} + S\Phi_{JKL}S\Phi_{IM} + S\Phi_{JKM}S\Phi_{IL} + S\Phi_{JLM}S\Phi_{IK} + S\Phi_{KLM}S\Phi_{IJ} \\ +S\Phi_{IJKL}S\Phi_{M} + S\Phi_{IJKM}S\Phi_{L} + S\Phi_{IJLM}S\Phi_{K} + S\Phi_{IKLM}S\Phi_{J} + S\Phi_{JKLM}S\Phi_{I} \end{bmatrix}$$

$$-\frac{d^3F}{d\rho^3}\begin{bmatrix} S\Phi_{IJK}S\Phi_{L}S\Phi_{M} + S\Phi_{IJL}S\Phi_{K}S\Phi_{M} + S\Phi_{IJM}S\Phi_{K}S\Phi_{L} + S\Phi_{IKL}S\Phi_{J}S\Phi_{M} \\ +S\Phi_{IKM}S\Phi_{J}S\Phi_{L} + S\Phi_{ILM}S\Phi_{J}S\Phi_{K} + S\Phi_{JKL}S\Phi_{I}S\Phi_{M} + S\Phi_{JKM}S\Phi_{I}S\Phi_{L} \\ +S\Phi_{JLM}S\Phi_{I}S\Phi_{K} + S\Phi_{KLM}S\Phi_{I}S\Phi_{J} \\ +S\Phi_{IJ}S\Phi_{KL}S\Phi_{M} + S\Phi_{IJ}S\Phi_{KM}S\Phi_{L} + S\Phi_{IJ}S\Phi_{LM}S\Phi_{K} + S\Phi_{IK}S\Phi_{JL}S\Phi_{M} \\ +S\Phi_{IK}S\Phi_{JM}S\Phi_{L} + S\Phi_{IK}S\Phi_{LM}S\Phi_{J} + S\Phi_{IL}S\Phi_{JK}S\Phi_{M} + S\Phi_{IL}S\Phi_{JM}S\Phi_{K} \\ +S\Phi_{IL}S\Phi_{KM}S\Phi_{J} + S\Phi_{IM}S\Phi_{JK}S\Phi_{L} + S\Phi_{IM}S\Phi_{JL}S\Phi_{K} + S\Phi_{IM}S\Phi_{KL}S\Phi_{J} \\ +S\Phi_{JK}S\Phi_{LM}S\Phi_{I} + S\Phi_{JL}S\Phi_{KM}S\Phi_{I} + S\Phi_{JM}S\Phi_{KL}S\Phi_{I} \end{bmatrix}$$

$$-\frac{d^4F}{d\rho^4}\begin{bmatrix} S\Phi_{IJ}S\Phi_{K}S\Phi_{L}S\Phi_{M} + S\Phi_{IK}S\Phi_{J}S\Phi_{L}S\Phi_{M} + S\Phi_{IL}S\Phi_{J}S\Phi_{K}S\Phi_{M} \\ +S\Phi_{IM}S\Phi_{J}S\Phi_{K}S\Phi_{L} + S\Phi_{JK}S\Phi_{I}S\Phi_{L}S\Phi_{M} + S\Phi_{JL}S\Phi_{I}S\Phi_{K}S\Phi_{M} \\ +S\Phi_{JM}S\Phi_{I}S\Phi_{K}S\Phi_{L} + S\Phi_{KL}S\Phi_{I}S\Phi_{J}S\Phi_{M} + S\Phi_{KM}S\Phi_{I}S\Phi_{J}S\Phi_{L} \\ +S\Phi_{LM}S\Phi_{I}S\Phi_{J}S\Phi_{K} \end{bmatrix}$$

$$-\frac{d^5F}{d\rho^5} \; S\Phi_{I}S\Phi_{J}S\Phi_{K}S\Phi_{L}S\Phi_{M}$$

(A10)

# Appendix B: Parameters of the interatomic interactions under examination for pure copper

After a short description of the basic ingredients entering the definition of each potential, the resulting elastic constants (second, third and fourth order) are gathered in table B2, the formation and migration energies, volumes and entropies in table B3 at the end of the Appendix. Only those remarks specific for each potential are kept within the following subsections.

For all potentials, the tri-vacancy energy barrier has a double hump. In the starting configuration, the jumping atom is located on (0,0,0) and the tri-vacancy on (1,1,0) + (0,1,1)+(1,0,1); the jump exchanges the atom and the vacancy initially located on (1,1,0). The saddle is found along the <111> direction. An intermediate metastable position shows up, the jumping atom being located at (1/2,1/2,1/2). The depth of the well of metastability depends sharply on the range of the potentials; it corresponds to a shallow minimum for most potentials but one. The completion of the jump is performed over a second barrier of the same height in the <11 –1> direction.

## B.1. Finnis-Sinclair type with exponential damping and cutoff at finite distance (FS-ED-2.82)

The expressions of the nude interactions are damped by an exponential which ensures that the functions and all their derivatives at any order vanish at the cutoff radius.

$$V(r_{ij}) = \alpha \exp\{-p\, r_{ij}/r_o\} \; \exp\{a_{pair}/(r_{cut} - r_{ij})\}$$

$$\Phi(r_{ij}) = \beta \exp\{-2q\, r_{ij}/r_o\} \; \exp\{a_{phi}/(r_{cut} - r_{ij})\} \quad (B1)$$

$$\rho_i = \sum_{j \neq i} \Phi(r_{ij}) \qquad Fnbod(\rho_i) = \rho_i^{1/2}$$

The cutoff distance is chosen in order to yield approximately the same values of $V(r)$ and $\Phi(r)$ at first, second and third neighbour distances as those of the nude interaction which was used in FS-PT345-2.00 and FS-PT456-2.00; the coefficients $a_{pair}$ and $a_{phi}$ of the damping functions are chosen to keep a positive curvature all along the curve. Preliminary graphic examination shows that the seventh neighbour shell must be included, even if its numerical contribution is very small. The corresponding coefficients are listed hereafter:

$\alpha = 0.2356438$ eV $\qquad p = 10.26788 \qquad a_{pair} = -1$ Angs

$\beta = 2.476014$ eV$^2$ $\qquad q = 2.226221 \qquad a_{phi} = -2$ Angs

$r_o = 2.556191$ Angs $\qquad r_{cut} = 2.81957444\; r_o$

For the tri-vacancy jump, the top of the barrier is located at (0.347, 0.347, 0.347); the depth of the local metastable minimum is -0.008 eV.

## B.2. Finnis-Sinclair type with exponential damping and no cutoff (FS-Barreteau)

This potential was primarily designed to study vibrational and thermodynamical properties of Cu surfaces. The interaction range is kept small enough to match specific physical properties of copper surfaces, namely a correct sign and magnitude for the surface relaxation and high frequency (optical like) vibrational modes at surfaces clearly separated from those of the bulk by a gap of the right order of magnitude : it retains mainly the contribution of first and second neighbours through the use of a damping function which is tuned to decay exponentially beyond a threshold radius $r_c$ slightly larger than the second neighbour distance.

$$V(r_{ij}) = \alpha \ (r_o/r_{ij})^p \quad \left\{1+\exp\left\{\frac{r_{ij}-r_c}{\delta}\right\}\right\}^{-1}$$

$$\Phi(r_{ij}) = \exp\left\{-2q \ r_{ij}/r_o\right\} \quad \left\{1+\exp\left\{\frac{r_{ij}-r_c}{\delta}\right\}\right\}^{-1} \quad (B2)$$

$$\rho_i = \sum_{j \neq i} \Phi(r_{ij}) \qquad \text{Fnbod}(\rho_i) = \beta \ \rho_i^{2/3}$$

The damping function fc is equal to 0.9997, $4.1 \ 10^{-4}$ and $4.1 \ 10^{-10}$ at second, third and fourth neighbour distance respectively. Although such a formulation would require in principle no cutoff radius, we used one set equal to 2.25 $r_o$ between fourth and fifth neighbours in order to avoid an artificially null elastic constant $C_{456}$ as mentionned in [28] and we checked that this modification did not bring any detectable change in the energies or entropies.

The corresponding coefficients are listed below:

$\alpha = 0.41112644734228$ eV      $p = 7.2055121109404$

$\beta = 1.1021047271284$ eV      $q = 2.2205754785209$

$r_o = 2.55265548$ Angs      $r_c = 1.5748307 \ r_o$      $\delta = 0.05$

For tri-vacancy, the saddle point configurations raised difficulties; although the final state was relaxed correctly (residual force components as low as $10^{-5}$ eV Angs$^{-1}$), several imaginary frequencies were found instead of one as expected.

The interaction range of this potential is so short that for the trivacancy jump, the local minimum in the intermediate metastable position becomes more stable than the stable starting configuration. The barrier top is somewhere around (0.25, 0.25, 0.25). For the dumbbell, no reasonable saddle configuration could even be found with the drag method and more sophisticated algorithms should be used.

**B.3. EAM-Mishin**

**B.3.1. Original EAM-Mishin.** The original version of this potential has a pair term close to a Morse-function which has a shallow minimum around the equilibrium distance $r_o$. The density function is monotously decreasing and the damping function for the two contributions is a rational fraction. Additional short ranged terms are used to harden the repulsion at close approach distances:

$$V(r_{ij}) = \left[ E_1 M(r_{ij}, r_O^{(1)}, \alpha_1) + E_2 M(r_{ij}, r_O^{(2)}, \alpha_2) + \delta \right] fc\left( \frac{r_{ij} - r_{cut}}{h} \right)$$

– short ranged terms

$$\Phi(r_{ij}) = \left[ a \exp\left\{ -\beta_1 (r_{ij} - r_O^{(3)})^2 \right\} + \exp\left\{ -\beta_2 (r_{ij} - r_O^{(4)}) \right\} \right] fc\left( \frac{r_{ij} - r_{cut}}{h} \right) \quad (B3)$$

$$M(r, r_O, \alpha) = \exp\left\{ 2\alpha(r - r_O) \right\} - 2 \exp\left\{ \alpha(r - r_O) \right\}$$

$$fc(x) = \frac{x^4}{1 + x^4}$$

$$\rho_i = \sum_{j \neq i} \Phi(r_{ij})$$

The embedding function is represented by :

$$F(\rho) = F_o + \frac{1}{2} F_2 (\rho - \rho_o)^2 + \sum_{n=1}^{4} q_n (\rho - \rho_o)^{n+2}$$

$$F(\rho) = \frac{F_o + \frac{1}{2} F_2 (\rho - \rho_o)^2 + q_1 (\rho - \rho_o)^3 + Q_1 (\rho - \rho_o)^4}{1 + Q_2 (\rho - \rho_o)^3} \quad (B4)$$

for $\rho < \rho_O$ and $\rho > \rho_O$ respectively. The density $\rho_O$ is further normalised to unity. Choosing for the embedding function different analytical formulas around the equilibrium value $\rho_O$ of the density leads to a discontinuity of its third order derivative. The latter is clearly evidenced when plotting the second order elastic constants $C_{11}$, $C_{12}$ and the bulk modulus B versus lattice parameter 'a': all three of them decrease when 'a' increases and they exhibit a slight cusp around the equilibrium value $a_o$ (unlike $C_{44}$ which doesn't). As a consequence, the calculation of entropies of a defective configuration (vacancy, dumbbell, saddle configuration) might be altered by this ill-placed discontinuity because all these configurations involve a probing of $F(\rho)$ on the two sides of the discontinuity.

It is easily checked, however, that in the present case the consequences are not too deleterious. We compared the values of the formation entropies for vacancy and dumbbell defect with those obtained if the analytical formula of the embedding function designed for $\rho < \rho_O$ is extended above $\rho_O$ (case denoted by 'low-rho') or, conversely, if the analytical formula designed for $\rho < \rho_O$ is extended below $\rho_O$ ('high-rho'). Fortunately the differences are negligible for the energies and for the activation volumes; for the entropies, when noticeable, they remain small enough to be neglected as can be seen in table B1.

**B.3.2. Modified EAM-Mishin**

The embedding function must be represented by a function having none of the drawbacks enlightened above. It is possible to reproduce the results already published for the original version of this potential through a simple fitting of the N-body function by a fifth order polynomial for some neighbourhood of $\rho_O$. For defect calculations (formation or migration), it can be checked that the range of $\rho$ values which is probed is rather narrow $0.5 < \rho/\rho_O < 1.5$. As a consequence, a polynomial function of the form

$$F(\rho) = q_o + \sum_{i=1}^{4} q_i (\rho - \rho_o)^{i+1} \quad (B5)$$

was fitted to the original values of the embedding function (available at http://cst-www.nrl.navy.mil/ccm6/ap/eam/index.html) in this interval and the coefficients were found to be :

$q_o = -2.2823539 \quad q_1 = 0.733148257 \quad q_2 = -0.763132346$
$q_3 = 0.124172653 \quad q_4 = 0.683385904$

This modified analytical form of the embedding function was used to calculate the entropy change under strain. The basic quantities pertaining to defect formation and migration are gathered in table B1 and are very close to the ones published for the original version.

For the tri-vacancy jump, the top of the barrier is located at (0.327, 0.327, 0.327); the depth of the local metastable minimum is -0.025 eV.

(insert table B1 around here)

**B.4. EAM-Milstein**

The potential was primarily designed to account for second and third order elastic constants. For that purpose, the equation of state (EOS) is modified by replacing, in some range around the equilibrium lattice parameter $a_o$, Rose's original expression by a seventh order polynomial:

$$EOS(a) = \begin{cases} -E_{coh} + \omega_1(a/a_o - 1)^2 + \omega_2(a/a_o - 1)^3 + \sum_{i=1}^{4} \gamma_i(a/a_o - 1)^{i+3} & \text{for } 0.95 < a/a_o \leq 1 \\ -E_{coh} + \omega_1(a/a_o - 1)^2 + \omega_2(a/a_o - 1)^3 + \sum_{i=1}^{4} \eta_i(a/a_o - 1)^{i+3} & \text{for } 1 \leq a/a_o < 1.05 \\ -E_{coh}(1 + a^* + k\,a^{*3})\exp\{-a^*\} & \text{for } |a/a_o - 1| > 0.05 \end{cases}$$

with $a^* = (a/a_O - 1)\left[(9\Omega_O B)/E_{coh}\right]^{1/2}$ where $\Omega_o$, B and $E_{coh}$ stand for the atomic volume, the bulk compressibility and the cohesive energy respectively.

The pair part of the potential is a sixth order polynomial with a polynomial decay at the cutoff distance; it exhibits a very shallow minimum around the first neighbour distance. The density function is oscillatory with no cutoff thanks to the high value of the power β in the denominator:

$$V(r_{ij}) = A \sum_{k=0}^{6} d_k (r_{ij})^k (r_{ij} - r_{cut})^4$$

$$\Phi(r_{ij}) = \frac{1 + b_1 \cos(\alpha\, r_{ij}) + b_2 \sin(\alpha\, r_{ij})}{(r_{ij})^\beta} . \quad (B6)$$

$$\rho_i = \sum_{j \neq i} \Phi(r_{ij})$$

The values of all the coefficients can be found in the original paper, taking due account of the fact that $\gamma_i$, $\eta_i$ and A must be divided by 1.60219. The EOS function is continuous up to third order only. No functional form for F(ρ) was proposed in the original paper and we explain hereafter the way how we designed one.

The values of ρ and F(ρ) were collected by varying the lattice parameter in the range [0.8 $a_o$ : 1.5 $a_o$]. The value of ρ at $a_o$ is denoted by $\rho_o$ and used later on to normalize the original expression of the density. The resulting normalised values of ρ belong to the interval [0.3 : 15]. In this range, the embedding function F(ρ) can be represented fairly well by an expression of the form:

$$F(\rho) = \exp\{P_{jp-, jp+}\} - \exp\{Q_{jq-, jq+}\} \quad (B7)$$

where $P_{jp-,jp+}$ and $P_{jq-,jq+}$ are polynomials containing powers of $\rho$ ranging from $\rho^{jp-}$ to $\rho^{jp+}$ and from $\rho^{jq-}$ to $\rho^{jq+}$ respectively, where jp- and jq- are negative, jp+ and jq+ are positive:

$$P = \sum_{jp-}^{jp+} p(i)\, r^i \qquad Q = \sum_{jq-}^{jq+} q(i)\, r^i . \qquad (B8)$$

The coefficients are then determined by a least square fit of the differences between the trial function and the target values. Numerous fits were tried and the best ones contained an equal number of positive and negative powers. The optimum was found for $P_{-3:3}$ and $Q_{-3:3}$ the coefficients of which are given below :

p(-3) = − 1.33414425293426 $10^{-1}$     q(-3) = + 1.56864346323235 $10^{-2}$

p(-2) = + 1.74078181863972 $10^{-1}$     q(-2) = − 1.04088096867450 $10^{-1}$

p(-1) = − 2.04037766897627           q(-1) = + 1.14168394156699 $10^{-1}$

p(0)  = + 2.46163195696694           q(0)  = + 8.57530003401110 $10^{-1}$

p(1)  = − 8.58259043509680 $10^{-3}$     q(1)  = − 8.33014047222441 $10^{-2}$

p(2)  = − 4.87850283177439           q(2)  = + 1.47083060315825 $10^{-3}$

p(3)  = − 2.81065015825138 $10^{-1}$     q(3)  = + 4.97499743917570 $10^{-5}$

In some interval denoted by [$\rho_{min}$ : $\rho_{max}$] around $\rho_o$, the analytical representation is systematically too high (by $10^{-5}$ relative) and the minimum is not located exactly at $\rho_o = 1$. In this interval we subtracted a further exponential - exp(C) where C is also a polynomial containing powers of $\rho$ ranging from jc- to jc+, together with two more terms diverging to minus infinity at the bounds of the above interval :

$$C = cdiv(1)/(\rho_{min} - \rho) + cdiv(2)/(\rho - \rho_{max}) + \sum_{jc-}^{jc+} c(i)\, r^i \qquad (B9)$$

The coefficients where fitted to the differences between the target values and those attained by the approximation above, together with a further constraint on the first three derivatives at $\rho_o=1$ :

c(-3) = − 1.33414425293426 $10^{-1}$     $\rho_{min}$ = + 0.86283783196531

c(-2) = + 1.74078181863972 $10^{-1}$     $\rho_{max}$ = + 1.0607730618120

c(-1) = − 2.04037766897627           cdiv(1) = + 0.14009094343182 $10^{-1}$

c(0)  = + 2.46163195696694           cdiv(2) = + 0.25927328768594 $10^{-1}$

c(1)  = − 8.58259043509680 $10^{-3}$

c(2)  = − 4.87850283177439

c(3)  = − 2.81065015825138 $10^{-1}$

A last remark is worth mentionning: while tuning the polynomial C in order to reproduce at best the third order elastic constants, it was observed that $C_{144}$, $C_{155}$ and $C_{456}$ are much less sensitive to the detail of the potential than $C_{111}$, $C_{112}$ and $C_{123}$, the latter being the less well reproduced elastic constants.

The top of the energy barrier for the tri-vacancy jump is located at (0.43,0.43,0.43) and the depth of the metastable point is only –0.003 eV.

(insert table B2, B3 here)

### B.5. Remarks

A few trends can be extracted after examining the above tables.
   With regard to elastic constants:

* elastic constants of second and third order are reasonably similar; those produced by EAM-Milstein are by construction better than the others (except $C_{123}$ which could not be tuned at will with the expression of the embedding function adopted above);

* for fourth order elastic constants, the last column of table B2 does not contain anything but a rough evaluation based on an approximate form of the repulsive interaction and on a value of $C_{111}$ arbitrarily set equal to 12 eV Angs$^{-3}$ [41]. Surprisingly, the fourth order elastic constants produced by the three first potentials are definitely better than those produced by EAM-Milstein, although the latter was fitted with more refinement than the others to elastic properties.

With regard to parameters monitoring the formation of point defects:

* the potential FS-Barreteau is noticeably different from the others because of its very short range: the formation energies of vacancy defects are larger but the formation energy of the dumbbell is smaller; the reverse holds for the formation entropies;

* EAM-Milstein exhibits systematically higher formation entropies than all the other potentials for vacancy defects; but the reverse is true for the dumbbell.

With regard to parameters monitoring the migration of point defects:

* the results obtained by the four potentials are close to one another: the only exception is the migration energy of the tri-vacancy for EAM-Milstein which is between twice and three times larger than the others.

* migration volumes of divacancy and tri-vacancy are increasingly negative for all potentials.

* the pre-exponential factors do not differ from each other by more than one order of magnitude.

# Appendix C: Effet of an N-body term on the vibrational entropy under a pure shear strain

The numerical results described in section 2 show that a positive entropy change is most often obtained for homogeneous shear strains at constant volume; the only potential to exhibit a negative sign in some cases (but not all) is the EAM-Mishin-mod. No clear dependence on the type of shear strains could be extracted; it is worth to ask whether specific ingredients of the interaction have an influence on the sign of the final effect. We show below that the presence of a non linear embedding function may be at the root of an entropy decrease through the study of a very simple system made up by a single atom vibrating two-dimensionally in the potential well of four immobile nearest neighbours in a square lattice.

We choose hereafter a simple interaction for which the energy per atom 'i' is given by the general form:

$$U_i = \frac{1}{2}\sum_{j \neq i} \alpha V(r_{ij}) - \beta F(\rho_i)$$

$$F(\rho_i) = \rho_i^\lambda \qquad \rho_i = \sum_{j \neq i} \Phi(r_{ij})$$

(C1)

where parameters $\alpha$, $\beta$, $\lambda$ are positive and the interactions $V$ and $\Phi$ are to be chosen below. This simple and versatile formulation can switch from a pure pair potential ($\lambda = 1$) to a multibody one ($\lambda \neq 1$). For sake of simplicity and tractability, the interactions are supposed to vanish between first and second neigbors.

The definition of the equilibrium first neighbour distance $r_o$ implies the existence of some relation ship between the parameters describing the interaction, to be establihed for each particular form of $V$ and $\Phi$.

The potential energy change when the central atom is moved from its equilibrium position (0 , 0) to (x , y) is then calculated: neighbors 1 and 3 along Ox are forced to sit at $+a_x$ and $-a_x$ respectively, while neighbors 2 and 4 along Oy are forced to sit at $+a_y$ and $-a_y$ respectively. The distances of the central atom at (x,y) to its neighbours are given by:

$$r_1^2 = (x - a_x)^2 + y^2 \qquad r_2^2 = x^2 + (y - a_y)^2$$
$$r_3^2 = (x + a_x)^2 + y^2 \qquad r_4^2 = x^2 + (y + a_y)^2$$

(C2)

'$a_x$' and '$a_y$' can be different from $r_o$ in a distorted system and will be fixed later in order to match a desired strain mode, namely:

* a pure expansion amounting to $2\varepsilon$ with $a_x = r_o(1+\varepsilon) \qquad a_y = r_o(1+\varepsilon)$  (C3)

* a pure shear with $a_x = r_o(1+\varepsilon)(1-\varepsilon^2)^{1/2} \qquad a_y = r_o(1-\varepsilon)(1-\varepsilon^2)^{1/2}$  (C4)

The general expression of the energy is then:

$$U = \frac{1}{2}\sum_{j=1}^{4} \alpha V(r_j) - \beta \left[\sum_{j=1}^{4} \Phi(r_j)\right]^\lambda$$

$$+ \frac{1}{2}\alpha V(r_1) - \beta\left[\Phi(r_1) + \Phi(a_x) + 2\Phi(a_y)\right]^\lambda$$

$$+ \frac{1}{2}\alpha V(r_2) - \beta\left[\Phi(r_2) + \Phi(a_y) + 2\Phi(a_x)\right]^\lambda \qquad \text{(C5)}$$

$$+ \frac{1}{2}\alpha V(r_3) - \beta\left[\Phi(r_3) + \Phi(a_x) + 2\Phi(a_y)\right]^\lambda$$

$$+ \frac{1}{2}\alpha V(r_4) - \beta\left[\Phi(r_4) + \Phi(a_y) + 2\Phi(a_x)\right]^\lambda$$

The first term is the contribution of the central atom; the next 4 terms are the contributions of the four immobile neighbours embedded in their own surroundings.

For a distorted system $a_x$ and $a_y$ are no longer equal to $r_o$; but if their values do not depart too much from $r_o$, it is easy to show that the equilibrium position of the central atom is still given by $x = y = 0$.

The calculation of the second order derivatives at rest, in the presence of a strain $\varepsilon$, is straightforward. It is readily shown that only diagonal terms are non vanishing and that $\left.\frac{\partial^2 U}{\partial y^2}\right|_{eq}^{\varepsilon}$ can be deduced from $\left.\frac{\partial^2 U}{\partial x^2}\right|_{eq}^{\varepsilon}$ simply by interchanging '$a_x$' and '$a_y$'.

The determinant $\Delta'$ of equation 6 to be calculated in the distorted state is expressed as a function of the determinant $\Delta$ in the undistorted state under the form:

$$\Delta' = \left.\frac{\partial^2 U}{\partial x^2}\right|_{eq}^{\varepsilon} \left.\frac{\partial^2 U}{\partial y^2}\right|_{eq}^{\varepsilon} = \left.\frac{\partial^2 U}{\partial x^2}\right|_{eq}^{o} \left.\frac{\partial^2 U}{\partial y^2}\right|_{eq}^{o} [1 + 2\varepsilon\,\mu] = \Delta\,[1 + 2\varepsilon\,\mu] \qquad (C6)$$

for a pure dilatation and

$$\Delta' = \Delta\,[1 + \nu\,\varepsilon^2] \qquad (C7)$$

for a pure shear respectively.

The sign of the coefficients $\mu$ and $\nu$ determines the sign of the entropy change under strain: i) $\mu$ is expected to be negative since a positive volume expansion implies an increase of the entropy (equation 11) which implies in turn a decrease of the determinant (equation 6); ii) $\nu$ was observed numerically to be negative in most cases.

## C1. Exponentially decaying repulsion and density functions

The pair interaction, the density $\rho_i$ and the embedding function $F(\rho_i)$ are given by:

$$V(r_{ij}) = \exp\{-p\,r_{ij}/r_o\} \qquad \Phi(r_{ij}) = \exp\{-2q\,r_{ij}/r_o\} \qquad (C8)$$

Parameters p, q are positive. The definition of the equilibrium distance $r_o$ implies:

$$\alpha\,p\exp\{-p\} = 4^{\lambda}\beta q\lambda\,\exp\{-2q\lambda\} \qquad (C9)$$

As a condition of stability to be satisfied by this interaction, the value of the second derivative in the undistorted sytem must be positive

$$\left.\frac{\partial^2 U}{\partial x^2}\right|_{eq}^{0} = \frac{2\alpha p(p - 2q - q(\lambda - 1)/4)\exp\{-p\})}{r_o^2} \qquad (C10)$$

which requires $p - 2q - q(\lambda - 1)/4 > 0$. This condition is slightly more demanding than that required by a positive formation energy of the vacancy, namely $p - 2q > 0$ [42].

The final expression of $\mu$ is given by:

$$\mu = \frac{p - p^2 + \lambda(4q^2 - 2q) + \lambda(\lambda - 1)q^2/2}{p - 2q - q(\lambda - 1)/4} \qquad (C11)$$

A graphic examination of this function shows that it is always negative, provided the condition $p - 2q - q(\lambda - 1)/4 > 0$ is fulfilled.

In the case of a pairwise interaction ($\lambda=1$), the expression above reduces to:

$$\mu = -(p + 2q - 1) \qquad (C12)$$

which is always negative for commonly accepted values of p and q [42].

The final expression of $\nu$ is given by:

$$\nu = \frac{p^3 - 2p^2 - p - 8\lambda(q^3 - q^2) - 2q(\lambda - 2) - (\lambda - 1)q^2(4q + \lambda(2q - 1))/2}{p - 2q - q(\lambda - 1)/4}$$

$$- \left[ \frac{p^2 - 4q^2 + p - 2q - (\lambda - 1)q^2}{p - 2q - q(\lambda - 1)/4} \right]^2 \qquad (C13)$$

Graphic examination of this function shows that ν can take positive values for some range of the doublet (p,q) if λ is not equal to unity and produce negative entropy changes under pure shear strains.

For a pairwise interaction ($\lambda = 1$), the potential mimics Morse type functions and the coefficient reads:

$$\nu = -(2pq + 4p + 8q + 2) \qquad (C14)$$

and is always negative.

With the values of p, q, λ used for the potential FS-ED-2.82 given in Appendix B, it can be readily checked that a negative value for ν is obtained. And according to this prediction, although a much larger cutoff radius was used in the numerical simulations, only positive entropy changes were indeed found for pure shear strains.

### C2. Power law repulsion and exponentially decaying density functions

The potential FS-Barreteau mixes the two types of interaction:

$$V(r_{ij}) = (r_o / r_{ij})^p \qquad \Phi(r_{ij}) = \exp\{-2q\, r_{ij}/ r_o\} \qquad (C15)$$

Parameters p, q are positive. The definition of the equilibrium distance $r_o$ implies:

$$\alpha\, p = 4^\lambda \beta q \lambda\, \exp\{-2q\lambda\} \qquad (C16)$$

The final expressions of μ and ν are given by:

$$\mu = -\frac{p^2 + 2p - q^2[5\lambda - 1 + (\lambda - 1)(\lambda - 2)/2] + 2q\lambda + 1}{p - 2q + 1 - q(\lambda - 1)/4}$$

$$\nu = \frac{p^2 + 2p - q^3[8 + (\lambda + 2)(\lambda - 1)/2] + q^2[8 + \lambda(\lambda - 1)/4] - 2q(\lambda - 2) + 1}{p - 2q + 1 - q(\lambda - 1)/4} \qquad (C16)$$

$$- \left[ \frac{q^2(\lambda + 3) - p^2 + 2q - 4p - 3}{p - 2q + 1 - q(\lambda - 1)/4} \right]^2$$

With the values of p, q, λ of the potential FS-Barreteau given in Appendix B, it can be checked that negative values for μ and ν are obtained. And according to this prediction, although a slightly larger cutoff radius including also the second neighbours was used in our calculations, only positive entropy changes were found for pure shear strains.

### C3. Power law decaying repulsive and density functions

The same qualitative conclusions are found when replacing the exponential decaying functions by power laws mimicking Lennard-Jones type of potentials:

$$V(r_{ij}) = (r_o / r_{ij})^p \qquad \Phi(r_{ij}) = (r_o / r_{ij})^{2q} \qquad (C17)$$

The definition of $r_o$ implies now $\alpha p = 4^\lambda \beta q \lambda$. The coefficients μ, ν read :

$$\mu = -\frac{p^2 - 4\lambda q^2 + 2p - 4q - (\lambda-1)q(q\lambda+1)/2}{p - 2q - q(\lambda-1)/4}$$

$$\nu = \frac{(p-2q)(p^2 - 2pq + 4q^2 + 4p + 8q - (\lambda-1)(q^3(10+\lambda) + 4q^2 + q)}{p - 2q - q(\lambda-1)/4}$$
$$- \left[\frac{(p+2q+4)(2q-p) + p - 2q + (\lambda-1)q(2q+1)/2}{p - 2q - q(\lambda-1)/4}\right]^2 \qquad \text{(C13)}$$

A graphic examination shows that $\mu$ is negative for commonly chosen values of p and q. However $\nu$ can become positive for large values of $\lambda$.

In the case of a pairwise interaction, these coefficients reduce to $\mu = -(p + 2q + 2)$ and $\nu = -(2pq + 4p + 8q + 12)$; they are always negative and associated with positive entropy changes under pure shear strains.

## C4. Conclusion

The above remarks cannot be taken as a general proof : the interactions are specific, have a very limited range and a monotonous behaviour versus distance; moreover the repulsion and attraction are provided only by the pair and the N-body term respectively and the embedding function is a monotonic function of the density. Most of the available potentials which are used today are much more complicated : for EAM-Mishin-mod and EAM-Milstein, the pairwise interaction gives a small additional contribution to the attractive term and the embedding function is no longer monotonic but rather exhibits a minimum around the equilibrium density, thus preventing the above analysis from being straightforwardly applied. Nevertheless, one is left with the overall feeling that the presence of a multibody term ($\lambda \neq 1$) can change qualitatively the behaviour of the system and is able to switch the sign of the coefficient $\nu$ above from negative to positive.

**Table captions**

Table 1. Definition of the set of homogeneous strains applied to the simulation cell. Columns 1 to 6 give the independent components of the strains; column 7 gives the corresponding volume expansion; column 8 defines the numbers attached to the strains $\eta$ defined by **J** ; column 9 defines the numbers attached to the strains at constant volume defined by $\mathbf{J} \, D^{-1/3}$.

Table 2. Coefficients of the Finnis-Sinclair potential for copper with various termination polynomials and cutoff distances. $\alpha$ and $\beta$ are in eV, $r_o$ in Angs.

Table 3. Formation energy (in eV) and entropy (in units of k) of vacancy and dumbbell defect obtained by Finnis-Sinclair potentials for copper with various terminations.

Table 4. Entropic constants of first, second and third order obtained with the four potentials investigated. The constants are expressed in units of k, per atom and per unit strain.

Table 5. Entropic corrections (in units of k) obtained for the stable and saddle configurations of a vacancy jump with potential FS-ED-2.82. The expression of the energy embodies second, third or fourth order terms and the corresponding contributions due to dilatation and shears are displayed separately; the last line is the correction based on the only knowledge of the local pressure.

Table B1. Formation and migration energies, entropies, relaxation volumes for single vacancy (V), divacancy (V2), trivacancy (V3) and <100> dumbbell (I) defects in copper with potential EAM-Mishin in its original and modified forms. Energies are expressed in eV, relaxation and migration volumes in unit of atomic volume $\Omega_o$, entropies in units of k, pre-exponential factors $\nu_o$ for migration in $s^{-1}$.

Table B2. Elastic constants of second , third and fourth order obtained by lattice summations with the four potentials investigated (in eV Angs$^{-3}$). The last column contains the experimental results for second and third order constants and estimations of fourth order constants.

Table B3. Formation and migration energies, entropies, relaxation volumes for single vacancy (V1), divacancy (V2), trivacancy (V3) and <100> dumbbell (I) defects in copper obtained by the four potentials investigated. Energies are expressed in eV, relaxation and migration volumes in unit of atomic volume $\Omega_o$, entropies in units of k, pre-exponential factors $\nu_o$ for migration in $s^{-1}$.

$$\mathbf{J} = \begin{bmatrix} 1+d_1\varepsilon & d_6\varepsilon & d_5\varepsilon \\ d_6\varepsilon & 1+d_2\varepsilon & d_4\varepsilon \\ d_5\varepsilon & d_4\varepsilon & 1+d_3\varepsilon \end{bmatrix} \quad \text{Det}[\mathbf{J}] = D \quad \eta = \mathbf{J} \quad \eta = \mathbf{J}\ D^{-1/3}$$

| $d_1$ | $d_2$ | $d_3$ | $d_4$ | $d_5$ | $d_6$ | $\Delta V / V$ | n° | n° |
|---|---|---|---|---|---|---|---|---|
| 1 | 0 | 0 | 0 | 0 | 0 | $1+\varepsilon$ | 1 | 2 |
| 1 | 1 | 0 | 0 | 0 | 0 | $1+2\varepsilon+\varepsilon^2$ | 3 | 4 |
| 1 | -1 | 0 | 0 | 0 | 0 | $1-\varepsilon^2$ | 5 | 6 |
| 1 | 1 | 1 | 0 | 0 | 0 | $1+3\varepsilon+3\varepsilon^2+\varepsilon^3$ | 7 | |
| 1 | 1 | -2 | 0 | 0 | 0 | $1-3\varepsilon^2-2\varepsilon^3$ | 8 | 9 |
| 0 | 0 | 0 | 1 | 0 | 0 | $1-\varepsilon^2$ | | 10 |
| 0 | 0 | 0 | 1 | 1 | 0 | $1-2\varepsilon^2$ | | 11 |
| 0 | 0 | 0 | 1 | -1 | 0 | $1-2\varepsilon^2$ | | 12 |
| 0 | 0 | 0 | 1 | 1 | 1 | $1-3\varepsilon^2+2\varepsilon^3$ | | 13 |
| 0 | 0 | 0 | 1 | -1 | 1 | $1-3\varepsilon^2-2\varepsilon^3$ | | 14 |
| 0 | 1 | -1 | 1 | 0 | 0 | $1-2\varepsilon^2$ | | 15 |
| 0 | 1 | -1 | -1 | 0 | 0 | $1-2\varepsilon^2$ | | 16 |
| 0 | 1 | -1 | 0 | 0 | 1 | $1-2\varepsilon^2+\varepsilon^3$ | | 17 |
| 0 | 1 | -1 | 0 | 0 | -1 | $1-2\varepsilon^2+\varepsilon^3$ | | 18 |
| 0 | 1 | -1 | 1 | 1 | 0 | $1-3\varepsilon^2-\varepsilon^3$ | | 19 |
| 0 | 1 | -1 | 1 | -1 | | $1-3\varepsilon^2-\varepsilon^3$ | | 20 |
| 0 | 1 | -1 | 1 | 1 | 1 | $1-4\varepsilon^2+2\varepsilon^3$ | | 21 |
| 0 | 1 | -1 | 1 | 1 | -1 | $1-4\varepsilon^2+2\varepsilon^3$ | | 22 |
| 0 | 1 | -1 | 1 | -1 | 1 | $1-4\varepsilon^2+2\varepsilon^3$ | | 23 |
| 2 | -1 | -1 | 1 | 1 | 1 | $1-6\varepsilon^2+4\varepsilon^3$ | | 24 |
| 2 | -1 | -1 | -1 | 1 | 1 | $1-6\varepsilon^2+4\varepsilon^3$ | | 25 |
| 2 | -1 | -1 | 1 | -1 | 1 | $1-6\varepsilon^2+4\varepsilon^3$ | | 26 |
| 0 | 0 | 0 | 1 | 1 | -1 | $1-3\varepsilon^2-2\varepsilon^3$ | 27 | |
| 1 | 0 | 0 | 1 | 0 | 0 | $1+\varepsilon-\varepsilon^2-\varepsilon^3$ | 29 | 28 |
| 1 | 0 | 0 | 0 | 1 | 1 | $1+\varepsilon-2\varepsilon^2$ | 31 | 30 |
| -1 | 0 | 0 | 1 | 0 | 0 | $1-\varepsilon-\varepsilon^2+\varepsilon^3$ | 33 | 32 |
| -1 | 0 | 0 | 0 | -1 | 1 | $1-\varepsilon-2\varepsilon^2$ | 35 | 34 |

Table 1

|  | FS-PT345-2.00 | FS-PT345-1.99 | FS-PT456-2.00 | FS-PT456-1.99 |
|---|---|---|---|---|
| $\alpha$ | 0.1763567 | 0.1763567 | 0.1763567 | 0.1763567 |
| p | 10.65757 | 10.65757 | 10.65757 | 10.65757 |
| $\beta$ | 1.244282 | 1.244282 | 1.244282 | 1.244282 |
| q | 2.310711 | 2.310711 | 2.310711 | 2.310711 |
| $r_o$ | 2.556191 | 2.556191 | 2.556191 | 2.556191 |
| $r_{rac} / r_o$ | 1.825 | 1.825 | 1.825 | 1.825 |
| $r_{cut} / r_o$ | 2.00 | 1.99 | 2.00 | 1.99 |
| a | 0.0012802749 | 0.0016106633 | 0.0049588370 | 0.0065623258 |
| b | − 0.0036272078 | − 0.0048943035 | −0.015988059 | −0.0022606673 |
| c | 0.0032057778 | 0.0045564447 | 0.014302409 | 0.0021467482 |
| a' | 2.7095181 | 3.3003279 | 9.6366319 | 12.404355 |
| b' | − 8.5344988 | − 11.070402 | − 33.173496 | − 45.400343 |
| c' | 7.4475034 | 10.257176 | 30.268994 | 43.987101 |

Table 2

|  | FS-PT345-2.00 | FS-PT345-1.99 | FS-PT456-2.00 | FS-PT456-1.99 |
|---|---|---|---|---|
| $E_F^V$ | 1.249 | 1.250 | 1.2496 | 1.2498 |
| $S_F^V/k$ | 4.71 | 1.55 | 1.60 | 1.39 |
| $E_F^I$ | 2.959 | 2.980 | 2.997 | 2.954 |
| $S_F^I/k$ | 18.95 | 12.72 | 12.25 | 11.33 |

Table 3

| Entropic constants | | FS ED-2.82 | FS Barreteau | EAM Mishin-mod | EAM Milstein |
| --- | --- | --- | --- | --- | --- |
| full name | short name | | | | |
| $\lambda_1$ | $\lambda_1$ | + 6.570 | + 6.199 | + 5.365 | + 7.054 |
| $\lambda_{11}$ | $\lambda_2$ | + 3.982 | + 2.77 | − 3.99 | − 23.00 |
| $\lambda_{12}$ | $\lambda_3$ | − 1.676 | + 0.42 | − 9.67 | − 64.93 |
| $\lambda_{44}$ | $\lambda_4$ | + 13.19 | + 13.70 | + 7.35 | + 14.50 |
| $\lambda_{111}$ | $\lambda_5$ | + 0.85 | − 143 | +120 | + 631 |
| $\lambda_{112}$ | $\lambda_6$ | + 9.4 | − 62 | +191 | + 1743 |
| $\lambda_{123}$ | $\lambda_7$ | + 50 | − 18 | + 7 | + 3025 |
| $\lambda_{144}$ | $\lambda_8$ | + 42 | − 0.011 | + 20 | + 96 |
| $\lambda_{155}$ | $\lambda_9$ | − 31 | − 99 | +177 | + 165 |
| $\lambda_{456}$ | $\lambda_{10}$ | − 0.347 | − 88 | − 7.3 | − 92 |

Table 4

| Correction | stable configuration | | saddle configuration | |
|---|---|---|---|---|
| with full strain tensor | $\dfrac{\Delta S_{out}^{stable}(\text{dilat})}{k}$ | $\dfrac{\Delta S_{out}^{stable}(\text{shear})}{k}$ | $\dfrac{\Delta S_{out}^{saddle}(\text{dilat})}{k}$ | $\dfrac{\Delta S_{out}^{saddle}(\text{shear})}{k}$ |
| O(2) | -0.615 | +0.00882 | -0.398 | +0.169 |
| O(3) | -0.601 | +0.00886 | -0.115 | +0.180 |
| O(4) | -0.601 | +0.00886 | -0.122 | +0.176 |
| with local pressure | $\dfrac{\Delta S_{out}^{stable}(\text{press})}{k}$ | | $\dfrac{\Delta S_{out}^{saddle}(\text{press})}{k}$ | |
| | -0.605 | | -0.201 | |

Table 5

|  | Mishin original | Mishin low-rho | Mishin high-rho | Mishin modified |
|---|---|---|---|---|
| $E_F^V$ | 1.272 | 1.272 | 1.270 | 1.273 |
| $S_F^V / k$ | 1.404 | 1.387 | 1.463 | 1.408 |
| $\Delta V_F^V / \Omega_O$ | + 0.701 | + 0.701 | + 0.714 | + 0.700 |
| $E_M^V$ | 0.689 | 0.688 | 0.687 | 0.689 |
| $\Delta V_M^V / \Omega_O$ | + 0.107 | + 0.106 | + 0.110 | + 0.106 |
| $10^{-13} \nu_O^V$ | 0.7560 | 0.7649 | 0.7772 | 0.7665 |
| $E_F^{V2}$ | 2.398 | 2.398 | 2.391 | 2.401 |
| $S_F^{V2} / k$ | 2.661 | 2.681 | 2.832 | 2.734 |
| $\Delta V_F^{V2} / \Omega_O$ | + 1.383 | + 1.383 | + 1.411 | + 1.385 |
| $E_M^{V2}$ | 0.364 | 0.365 | 0.362 | 0.364 |
| $\Delta V_M^{V2} / \Omega_O$ | - 0.130 | - 0.131 | - 0.135 | - 0.132 |
| $10^{-13} \nu_O^{V2}$ | 1.317 | 1.331 | 1.289 | 1.298 |
| $E_F^{V3}$ | 3.376 | 3.376 | 3.362 | 3.379 |
| $S_F^{V3} / k$ | 3.821 | 3.849 | 4.049 | 3.925 |
| $\Delta V_F^{V3} / \Omega_O$ | + 2.079 | + 2.079 | + 2.118 | + 2.086 |
| $E_M^{V3}$ | 0.0625 | 0.0626 | 0.0660 | 0.0643 |
| $\Delta V_M^{V3} / \Omega_O$ | - 0.360 | - 0.360 | - 0.387 | - 0.364 |
| $10^{-13} \nu_O^{V3}$ | 0.7144 | 0.7202 | 0.7474 | 0.7409 |
| $E_F^I$ | 3.063 | 3.067 | 3.067 | 3.073 |
| $S_F^I / k$ | 7.429 | 7.412 | 7.518 | 7.428 |
| $\Delta V_R^I / \Omega_O$ | + 0.834 | + 0.834 | + 0.846 | + 0.825 |
| $E_M^I$ | 0.0976 | 0.0979 | 0.0977 | 0.0974 |
| $\Delta V_M^I / \Omega_O$ | + 0.0398 | + 0.0413 | + 0.0395 | + 0.0386 |
| $10^{-13} \nu_O^I$ | 0.1992 | 0.2007 | 0.1994 | 0.1982 |

Table B1

|  | FS ED-2.82 | FS Barreteau | EAM Mishin modified | EAM Milstein | Experiments |
| --- | --- | --- | --- | --- | --- |
| $C_{11}$ | + 1.058 | + 1.113 | + 1.084 | + 1.104 | + 1.0997 |
| $C_{12}$ | + 0.766 | + 0.785 | + 0.789 | + 0.785 | + 0.7798 |
| $C_{44}$ | + 0.478 | + 0.517 | + 0.476 | + 0.510 | + 0.5104 |
| $C_{111}$ | - 9.055 | - 8.752 | - 7.983 | - 11.024 | - 12.483 |
| $C_{112}$ | - 5.299 | - 5.197 | - 4.128 | - 6.157 | - 7.6146 |
| $C_{123}$ | - 0.3077 | - 0.349 | + 0.477 | - 1.663 | - 3.1207 |
| $C_{144}$ | - 0.234 | - 0.233 | - 0.283 | - 0.838 | - 0.8239 |
| $C_{155}$ | - 4.619 | - 4.557 | - 4.256 | - 4.400 | - 4.4002 |
| $C_{456}$ | + 0.170 | + 0.098 | + 0.113 | + 0.160 | + 0.1560 |
| $C_{1111}$ | + 73.970 | + 75.100 | + 47.554 | + 10.155 | ≈ + [84 :168] |
| $C_{1112}$ | + 38.841 | + 31.394 | + 17.342 | + 16.637 | ≈ + [42 : 84] |
| $C_{1122}$ | + 38.162 | + 34.342 | + 17.084 | + 12.868 | ≈ + [42 : 84] |
| $C_{1123}$ | + 0.632 | - 1.944 | - 7.736 | - 54.352 | ≈ 0 |
| $C_{1144}$ | + 0.141 | - 2.698 | - 0.203 | - 6.280 | ≈ 0 |
| $C_{1155}$ | + 36.113 | + 28.149 | + 25.920 | + 44.207 | ≈ + [42 : 84] |
| $C_{1244}$ | + 0.603 | - 2.204 | + 0.2621 | + 10.784 | ≈ 0 |
| $C_{1266}$ | + 36.240 | + 32.341 | + 24.875 | + 54.976 | ≈ + [42 : 84] |
| $C_{1456}$ | - 0.400 | - 3.152 | + 0.312 | - 9.239 | ≈ 0 |
| $C_{4444}$ | + 35.472 | + 31.140 | + 25.017 | + 25.188 | ≈ + [42 : 84] |
| $C_{4455}$ | + 0.012 | - 2.817 | + 0.666 | - 5.809 | ≈ 0 |

Table B2

|  | FS ED-2.82 | FS Barreteau | EAM Mishin mod | EAM Milstein |
| --- | --- | --- | --- | --- |
| $E_F^V$ | 1.203 | 1.513 | 1.273 | 1.305 |
| $S_F^V/k$ | 1.387 | 1.232 | 1.408 | 2.355 |
| $\Delta V_F^V/\Omega_O$ | + 0.742 | + 0.763 | + 0.700 | + 0.762 |
| $E_M^V$ | 0.672 | 0.547 | 0.689 | 0.682 |
| $\Delta V_M^V/\Omega_O$ | + 0.155 | + 0.0696 | +0.106 | + 0.101 |
| $10^{-13}\, \nu_O^V$ | 1.372 | 0.4288 | 0.7665 | 0.6109 |
| $E_F^{V2}$ | 2.270 | 2.881 | 2.401 | 2.440 |
| $S_F^{V2}/k$ | 2.768 | 2.371 | 2.734 | 4.577 |
| $\Delta V_F^{V2}/\Omega_O$ | + 1.479 | + 1.541 | + 1.385 | + 1.543 |
| $E_M^{V2}$ | 0.358 | 0.278 | 0.364 | 0.442 |
| $\Delta V_M^{V2}/\Omega_O$ | - 0.099 | - 0.223 | - 0.132 | - 0.221 |
| $10^{-13}\, \nu_O^{V2}$ | 4.185 | 2.158 | 1.298 | 1.738 |
| $E_F^{V3}$ | 3.198 | 4.273 | 3.379 | 3.392 |
| $S_F^{V3}/k$ | 4.071 | 3.283 | 3.925 | 6.392 |
| $\Delta V_F^{V3}/\Omega_O$ | + 2.223 | + 2.371 | + 2.086 | + 2.383 |
| $E_M^{V3}$ | 0.061 | 0.0915 | 0.0643 | 0.231 |
| $\Delta V_M^{V3}/\Omega_O$ | - 0.368 | - 0.560 | - 0.364 | - 0.708 |
| $10^{-13}\, \nu_O^{V3}$ | 0.7352 | ? | 0.7409 | 1.371 |
| $E_F^I$ | 3.034 | 2.616 | 3.073 | 3.188 |
| $S_F^I/k$ | 10.245 | 11.017 | 7.428 | 9.998 |
| $\Delta V_R^I/\Omega_O$ | + 0.997 | + 0.770 | + 0.825 | + 1.225 |
| $E_M^I$ | 0.084 | ? | 0.0974 | 0.0741 |
| $\Delta V_M^I/\Omega_O$ | + 0.0287 | ? | + 0.0386 | + 0.0213 |
| $10^{-13}\, \nu_O^I$ | 0.1501 | ? | 0.1982 | 0.04712 |

Table B3

**Figure captions**

Figure 1. Entropy change (in units of k) of a cubic box containing 864 atoms under a uniform shear stress of amplitude $\varepsilon$ in a log-log plot versus $\varepsilon^2$ for different potentials. From top to bottom: the empty circles (○) stand for potential FS-PT345-2.00, the empty squares (□) stand for potential FS-PT456-2.00, the empty diamonds (◊) stand for potential FS-PT345-1.99, the multiply signs (×) stand for potential FS-PT456-1.99. Two straight lines of slope 1/2 (top) and 1 (bottom) are displayed for comparison.

Figure 2. Formation entropy of mono-vacancy (in units of k) versus the inverse number $N^*$ of atoms in the inner region. (a) FS-ED-2.82; (b) FS-Barreteau; (c) EAM-Mishin-mod; (d) EAM-Milstein. The 'plus' symbols (+) stand for the embedded crystallite results; the 'multiply' symbols (×) for the elastic correction including pressure; the empty squares (□) for the elastic correction including dilatation and shears; the filled triangles (▲) for the supercell results.

Figure 3. Formation entropy of tri-vacancy (in units of k) versus the inverse number $N^*$ of atoms in the inner region. (a) FS-ED-2.82; (b) FS-Barreteau; (c) EAM-Mishin-mod; (d) EAM-Milstein. Same meaning of symbols as in figure 2.

Figure 4. Formation entropy of dumbbell (in units of k) versus the inverse number $N^*$ of atoms in the inner region. (a) FS-ED-2.82; (b) FS-Barreteau; (c) EAM-Mishin-mod; (d) EAM-Milstein. Same meaning of symbols as in figure 2.

Figure 5. Pre-exponential factor for the vacancy jump (in $s^{-1}$) versus the inverse number $N^*$ of atoms in the inner region obtained with EAM-Mishin-mod. (a) single vacancy; (b) dumbbell; (c) di-vacancy; (d) tri-vacancy. Same meaning of symbols as in figure 2.

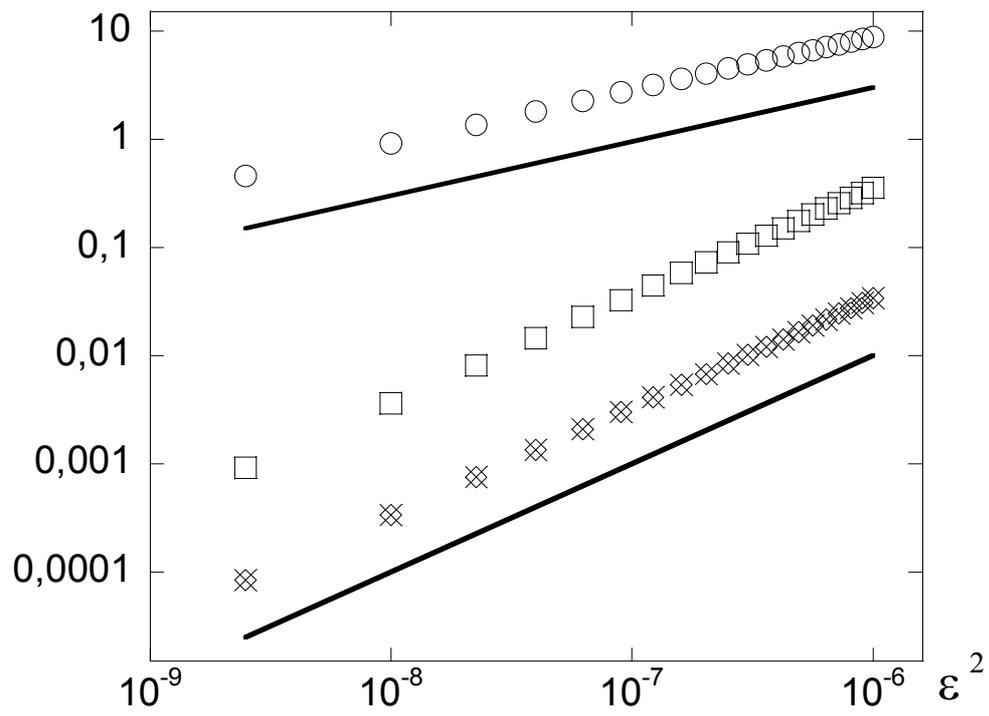

Figure 1

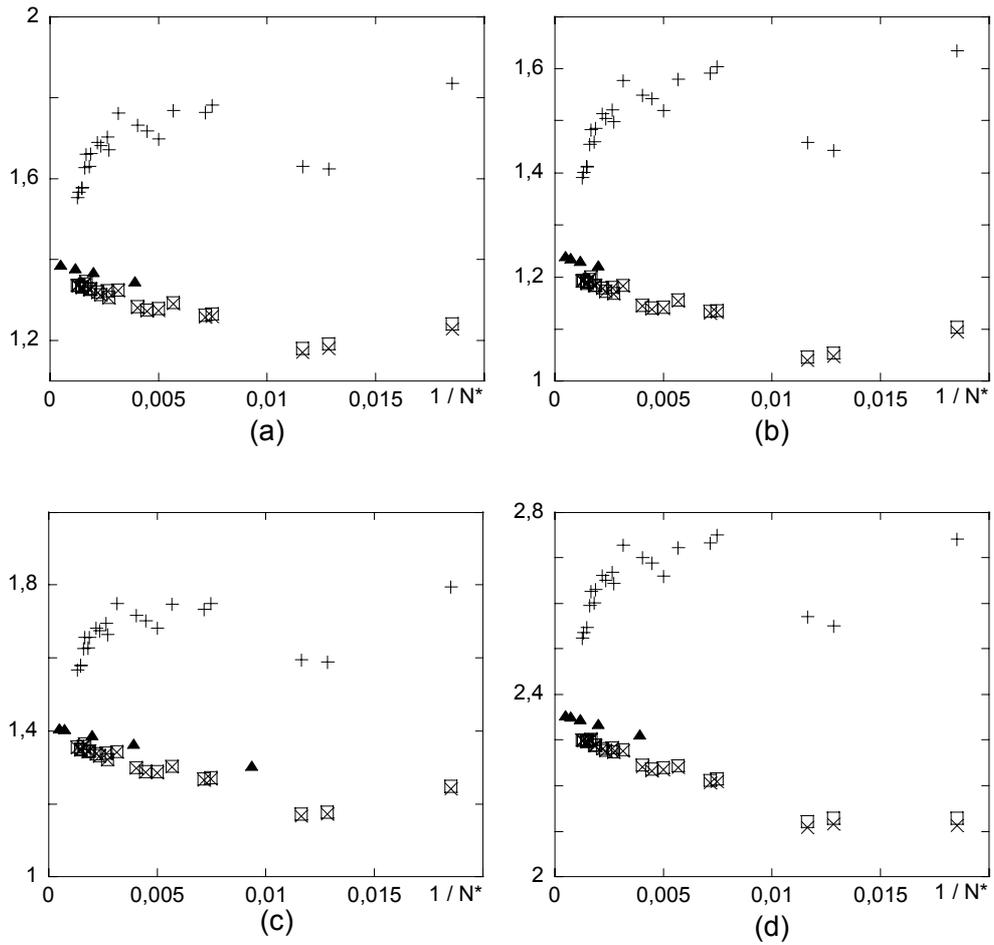

Figure 2

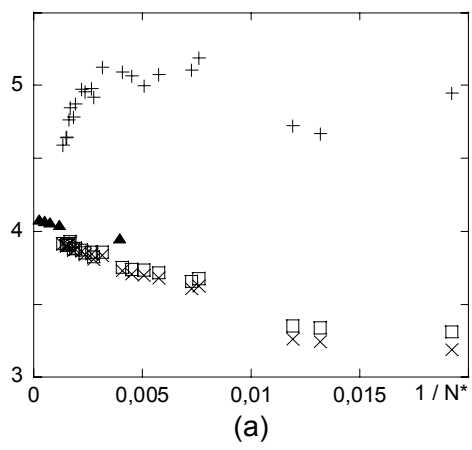
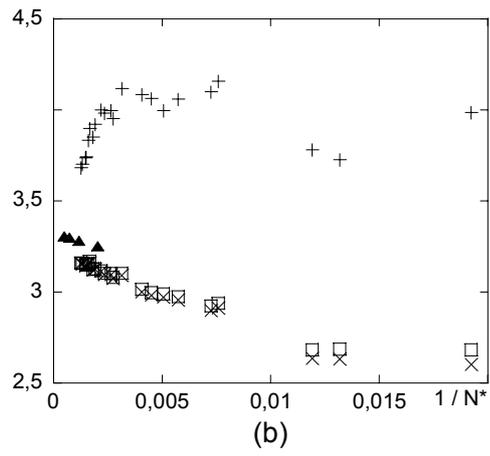
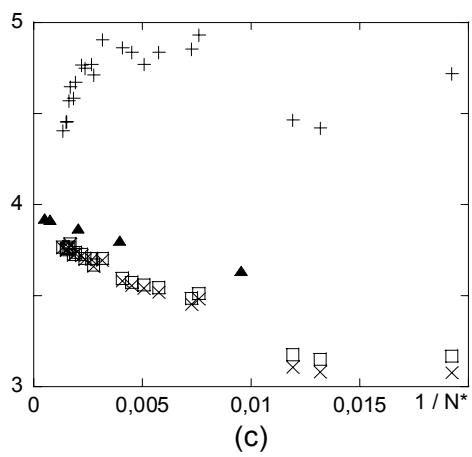
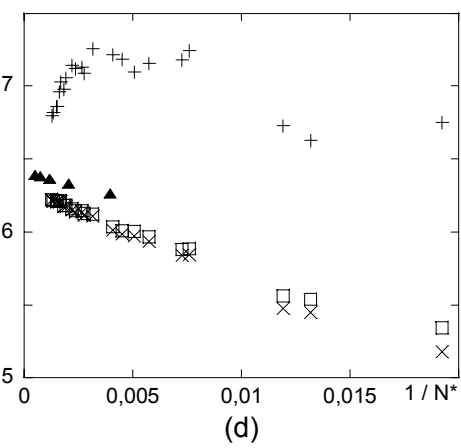

Figure 3

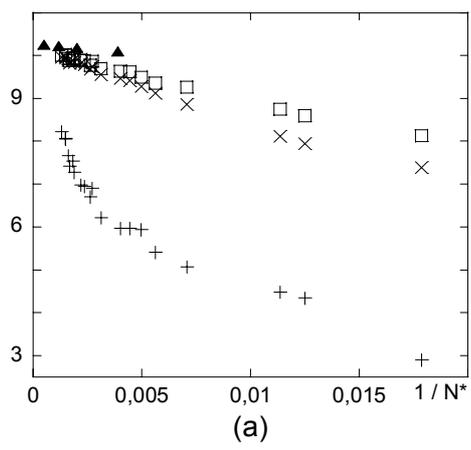
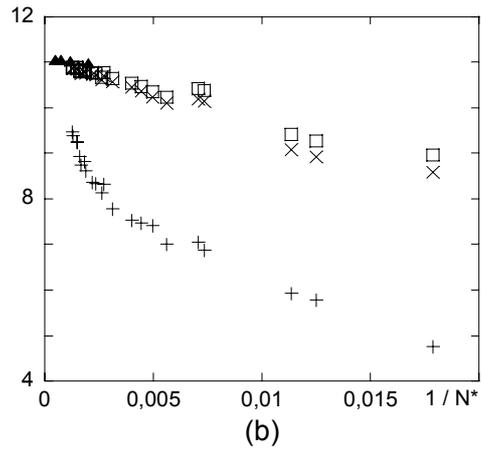
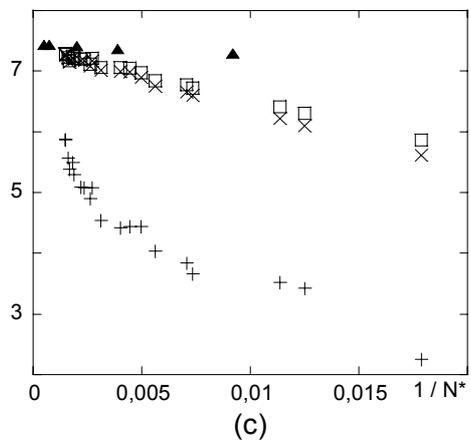
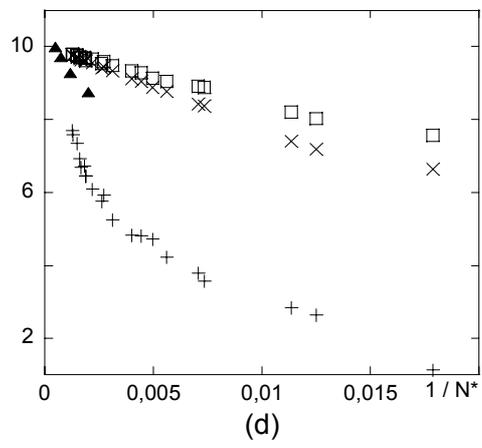

Figure 4

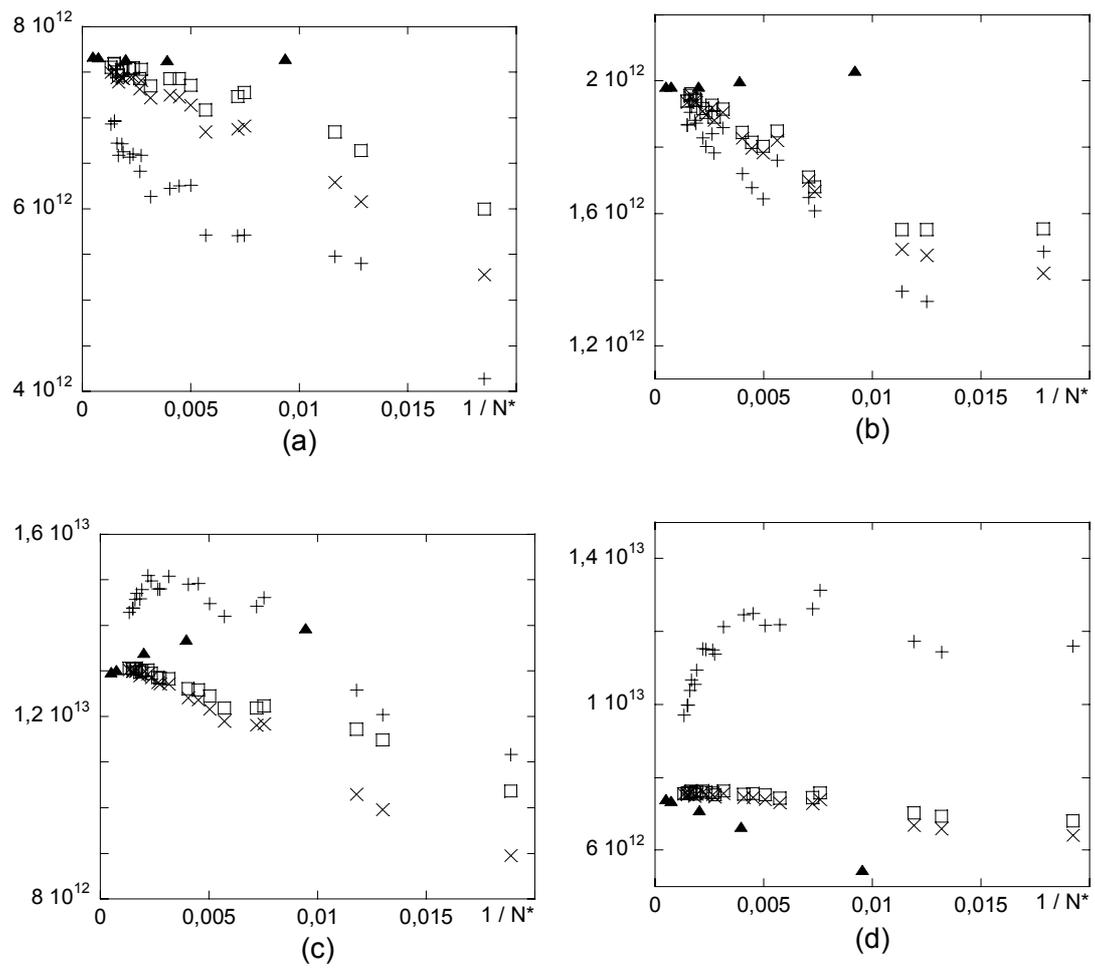

Figure 5